\documentclass[10pt]{article}

\usepackage{amsmath}
\usepackage{amssymb}

\usepackage{graphicx}

\usepackage{cite}

\usepackage{color} 

\usepackage{setspace} 
\usepackage[labelfont=bf,labelsep=period,justification=raggedright]{caption}

\makeatletter
\renewcommand{\@biblabel}[1]{\quad#1.}
\makeatother

\date{}

\makeatletter
\newcommand{\cn}{\mathop{\operator@font cn}}
\makeatother

\begin{document}

\begin{flushleft}
{\Large
\textbf{Production and Transfer of Energy and Information in Hamiltonian Systems}
}
\\
Ch. G. Antonopoulos$^{1,\ast}$, 
E. Bianco-Martinez$^{1}$, 
M. S. Baptista$^{1}$
\\
\bf{1} Department of Physics/ICSMB, University of Aberdeen, Aberdeen, United Kingdom
\\
$\ast$ E-mail: chris.antonopoulos@abdn.ac.uk
\end{flushleft}

\section*{Abstract}
We present novel results that relate energy and information transfer with sensitivity to initial conditions in chaotic multi-dimensional Hamiltonian systems. We show the relation among Kolmogorov-Sinai entropy, Lyapunov exponents, and upper bounds for the Mutual Information Rate calculated in the Hamiltonian phase space and on bi-dimensional subspaces. Our main result is that the net amount of transfer from kinetic to potential energy per unit of time is a power-law of the upper bound for the Mutual Information Rate between kinetic and potential energies, and also a power-law of the Kolmogorov-Sinai entropy. Therefore, transfer of energy is related with both transfer and production of information. However, the power-law nature of this relation means that a small increment of energy transferred leads to a relatively much larger increase of the information exchanged. Then, we propose an ``experimental'' implementation of a 1-dimensional communication channel based on a Hamiltonian system, and calculate the actual rate with which information is exchanged between the first and last particle of the channel. Finally, a relation between our results and important quantities of thermodynamics is presented.

\section*{Author Summary}

\section{Introduction}\label{Intro}

After the pioneering work by Shannon \cite{Shannon1948} on information, it became clear that it is a very useful and important concept as it can measure the amount of uncertainty an observer has about a random event and thus provides a measure of how unpredictable it is. The degree of disorder of a chaotic dynamical system is related to the degree of its chaotic behavior which is, in turn, characterized by the rate of exponential divergence of neighboring initial conditions, that is by the magnitude of the positive Lyapunov exponents \cite{Skokos2010}. It is the sensitive dependence on initial conditions \cite{Eckmann1985,Skokos2010,BouSkobook2012} that produces information since two different but indistinguishable initial conditions at a certain precision will evolve into distinguishable states after a finite time \cite{Eckmann1985}. This relation between production of information and sensitive dependence was made clear for systems that have absolutely continuous conditional measures \cite{Pesin1976,Pesin1977}, by:
\begin{equation}\label{HKS_entropy}
H_{\mbox{KS}}=\sum_{i}\lambda_i,
\end{equation}
where $H_{\mbox{KS}}$ represents the Kolmogorov-Sinai or KS entropy (Shannon's entropy per unit of time) and $\lambda_i$ are the positive Lyapunov exponents of the dynamical system \cite{Benettin1980a,Benettin1980b,Eckmann1985,Skokos2010}, which measure how sensitive to initial conditions the system is. This is a property that has been found to be true for many dynamical systems \cite{Eckmann1985}. In general, for bounded systems $H_{\mbox{KS}}\leqslant\sum_{i}\lambda_i,\;\lambda_i>0$ \cite{Ruelle1978}.

Energy and information can be produced in a system or transferred between its different ``parts'' or ``constituents'' \cite{Baptistaetal2012,Machiorietal2012,Mandaletal2013}. If transferred, there are always at least two ``entities'' involved. In general, they can be nodes, modes, or related functions that can be defined on subspaces or projections of the phase space of the system.

Another related concept to the Shannon entropy that can characterize random complex systems is the Mutual Information (MI) \cite{Shannon1948} which is a measure of how much uncertainty one has about a state variable after observing another state variable. For deterministic systems that present correlations, a more appropriate quantity for measuring the transfer of information is the Mutual Information Rate (MIR), MI per unit of time. In Refs. \cite{Baptistaetal2005,Baptistaetal2008A,Baptistaetal2008B,Baptistaetal2012}, the authors have developed alternative methods to overcome problems that stem from the definition of probabilities for these quantities and proposed the use of bounds for the MIR. In Ref. \cite{Baptistaetal2012}, the authors have derived an upper bound for the MIR between two nodes or two groups of nodes that depend on the largest Lyapunov exponents of the subspace of the network formed by the nodes. In particular, they have showed that:
\begin{equation}\label{Ic_MIR}
 \mbox{MIR}\leq I_c=l_1-l_2,\;l_1\geq l_2,
\end{equation}
where $l_1$ and $l_2$ are the two finite time and size Lyapunov exponents calculated in the bi-dimensional observation space which for simplicity will be referred herein as the Lyapunov exponents of the bi-dimensional subspace. In our study, when the observation space is formed by the kinetic ($K$) and potential ($P$) energy variables of the Hamiltonian, then the upper bound for  the MIR in the $KP$ space $I_c^{KP}=\lambda_1^{KP}-\lambda_2^{KP}$ (i.e. $l_1=\lambda_1^{KP}$ and $l_2=\lambda_2^{KP}$) represents the upper bound for the information transferred per unit of time between the kinetic and potential energies. The use of the $KP$ space to study the relationship between energy and information exchange is justifiable because the transfer of energy from kinetic to potential energy is easy and well understood. However, we will  also study this relationship in other bi-dimensional subspaces such as those formed by any two nodes of the Hamiltonian system.

The main result of our work is Eq. \eqref{first_result}, which states that  when considering specific energy subintervals, the time rate of energy transferred from the kinetic to the potential variable during a time step is a power-law function of either the largest Lyapunov exponent $\lambda_1$ of the Hamiltonian or of the upper bound $I_c^{KP}$ for the MIR of the bi-dimensional $KP$ space.

We then present the generalization of these power-law relations when considering much larger energy intervals of chaotic behavior with initial conditions set initially in different parts of the phase space of the same Hamiltonian system. We also consider different Hamiltonian systems in which we illustrate how they can be used to create communication systems.

The second main result is Eq. \eqref{second_result} which states that the upper bound for the MIR exchanged between the potential and kinetic energy is smaller than the upper bound for the MIR between two groups of oscillators formed each by half the oscillators of the Hamiltonian, and this is in turn smaller than the whole time rate of information produced by the system expressed by $H_{\mbox{KS}}$. We provide a proof of this result in the Appendix. This result implies that, when one observes a Hamiltonian system through its kinetic and potential energy (i.e. in its $KP$ space), one should not expect to have more information about the Hamiltonian system than when observing it directly (i.e. by observing half of its nodes or all of its variables).

The relation among energy, entropy, and information is a long lasting problem in physics. Nineteenth century saw the discovery of the two laws of thermodynamics, almost happening at the same time. The first law relates the rate of change of the energy of a body with the heat and work produced and the second, the rate of the change of the entropy of the body with the heating, implying the growth of its entropy during an adiabatic and irreversible process. Thermodynamics turned out to be a very important mathematical theory that can describe successfully macroscopic systems in equilibrium, based on the thermodynamic laws and provides a link between work, energy, and entropy as a universal competition, i.e. when a body approaches equilibrium, energy tends to a minimum and entropy to a maximum (see for example Ref. \cite{Muller2008}).

In 1929, after a long lasting controversy, Le\'o Szil\'ard \cite{Szilard1929} and more recently the authors in Refs. \cite{Landauer1961,Bennett1982}, showed that Maxwell's hypothetical demon does not contradict the second law of thermodynamics, implying that in principle one can convert information to free energy. By free energy we mean the portion of the energy of a system that is available to perform work mediated by thermal energy. It was only very recently in 2010 \cite{Toyabeetal2010}, that an experimental demonstration of this information to energy conversion has been achieved.

In Ref. \cite{Machiorietal2012}, the authors study the energy transfer in terms of the classical dynamics of two particles that move in harmonic potential wells, interacting with the same external environment of $N$ noninteracting chaotic systems. They found that the oscillators can exchange energy through the environment when in almost-perfect resonance and in Ref. \cite{Mandaletal2013}, a simple and solvable model of a device that transfer energy from a cold to a hot system by rectifying thermal fluctuations is presented. In order for this to happen, the device requires a memory register to which it can write information. The subtle issue of the connection between work and information processing is presented in Ref. \cite{Mandaletal2012} in a solvable model of an autonomous Maxwell's demon. The authors studied and explained a device that makes measurements about the system states, stores this information into a register, and delivers work by rectifying thermal fluctuations.

In this work however, we are interested in providing the relation between energy transfer and information production and transfer in multi-dimensional chaotic Hamiltonian systems, e.g. in isolated systems where the total energy of the system remains constant and no exchange of heat or matter with the surroundings exists. Such a relation could allow one to realize how much information a sort of Maxwell's demon would need in order to be able to transfer a certain amount of energy between oscillatory modes in Hamiltonian systems. Hamiltonian systems such as those we study herein differ from thermodynamic systems in the sense they are far from the thermodynamic limit, i.e. they have a small dimensionality. However, in the Discussion session, we provide a link between our results and important quantities of thermodynamics.

The paper is organized as follows: In Sec. \ref{Methods} we present the basic material needed in our study. This includes the presentation of the two Hamiltonian systems and some of its important properties, the definition of the $KP$ bi-dimensional observation space and a brief discussion about important quantities from the theory of information such as upper bound for MIR and KS entropy. In Subsec. \ref{section_Results_2013_02_12_A}, we present the relation between the largest Lyapunov exponent of the Hamiltonian system and that of the bi-dimensional space of the kinetic and potential energy. Then, in Subsec. \ref{section_Results_2013_02_12_B}, we explain how one can arrive at Eq. \eqref{first_result} about the relation between production and transfer of information when considering specific energy subintervals of chaotic behavior. In Subsec. \ref{large_nrg_regime_sec} we generalize the main results of our study by considering the case of different Hamiltonian systems for much larger energy intervals and with initial conditions set in different parts of the phase space of the systems. Then, in Sec. \ref{Ham_com_chaneel_FPU} we illustrate how one can implement a 1-dimensional communication channel based on a Hamiltonian system, and calculate the actual rate with which information is exchanged between the first and last particle of the channel. In the Discussion section we briefly recall the main results of our study, their implications and relation with quantities of thermodynamics. Finally, in the Appendix we provide a proof of the inequality presented in Eq. \eqref{second_result}.

\section{Materials and Methods}\label{Methods}
\subsection{Fermi-Pasta-Ulam Hamiltonian}\label{FPU_section}

In this work we use two different Hamiltonian systems. We first consider the 1-dimensional lattice of $N$ particles with equal masses and nearest neighbour interactions with quartic nonlinearities ($\beta$-model) given by the Fermi-Pasta-Ulam (FPU) Hamiltonian \cite{Fermietal1955}:
\begin{equation}\label{FPU_Hamiltonian_beta}
H(x,\dot{x})=\frac{1}{2}\sum_{j=1}^{N}\dot{x}_{j}^{2}+\sum_{j=0}^{N}\biggl
(\frac{1}{2}(x_{j+1}-x_{j})^2+\frac{1}{4}\beta(x_{j+1}-x_{j})^4\biggr)=E
\end{equation}
adopting fixed boundary conditions:
\begin{equation}\nonumber
x_{0}(t)=x_{N+1}(t)=0,\forall t.
\end{equation}
Here, $x_{j}$ and $\dot{x}_{j}$ is the position and conjugate momentum of the $j$th particle, respectively.

For this system, we use initial conditions in the neighborhood of two particular simple periodic orbits of \eqref{FPU_Hamiltonian_beta} which are called SPO1 and SPO2 \cite{Antonopoulosetal2006a,Antonopoulosetal2006b}. The reason for this choice is that they allow us to control in a systematic way the increase of the energy of the system so that chaotic motion will be sustained. Any other way of increasing the energy of the system so that chaotic behavior can exist may be equally used as well.

SPO1 is specified by the conditions:
\begin{equation}\label{FPU_non_lin_mode_fixed_boundary_conditions_SPO1}
x_{2j}(t)=0,\;x_{2j-1}(t)=-x_{2j+1}(t)\equiv\hat{x}(t),\;j=1,\ldots,\frac{N-1}{2},
\end{equation}
and exists for all odd $N$, keeping every even particle stationary
at all times. It is not difficult to show that this is, in fact, the
$q=(N+1)/2$ mode of the linear lattice (i.e. $\beta=0$) with frequency
$\omega_q=\sqrt{2}$. The remarkable property of this solution is
that it is continued in precisely the same spatial configuration in
the nonlinear lattice as well, due to the form of the equations of
motion associated with Hamiltonian (\ref{FPU_Hamiltonian_beta}):
\begin{equation}\label{FPU_eq_motion_SPO1}
\ddot{x}_{j}(t)=x_{j+1}-2x_{j}+x_{j-1}+\beta\Bigl((x_{j+1}-x_{j})^3-(x_{j}-x_{j-1})^3\Bigr),\;j=1,\ldots,N
\end{equation}
which reduce, upon using
(\ref{FPU_non_lin_mode_fixed_boundary_conditions_SPO1}) with
the fixed boundary conditions to a single second order
nonlinear differential equation for $\hat{x}(t)$:
\begin{equation}\label{FPU_single_equation_SPO1}
\ddot{\hat{x}}(t)=-2\hat{x}(t)-2\beta\hat{x}^{3}(t)
\end{equation}
describing the oscillations of all moving particles of SPO1, with
$j=1,3,5,\ldots,N$. For the stationary particles
(i.e. $j=2,4,6,\ldots,N-1$) we have $\hat{x}(t)=0,\forall
t\ge0$. The solution of (\ref{FPU_single_equation_SPO1}) is well
known in terms of Jacobi elliptic functions \cite{Abramowitzetal1965} and can be written as:
\begin{equation}\label{sol_FPU_single_equation_SPO1}
\hat{x}(t)=\mathcal{C}\cn(\lambda t,{\kappa}^{2}),
\end{equation}
where:
\begin{equation}\label{FPU_C_and_lambda_SPO1}
\mathcal{C}^{2}=\frac{2{\kappa}^{2}}{\beta(1-2{\kappa}^{2})},\
\lambda^{2}= \frac{2}{1-2{\kappa}^{2}},
\end{equation}
and ${\kappa}^{2}$ is the modulus of the $\cn$ elliptic function.
The energy per particle of SPO1 is then found to be:
\begin{equation}\label{FPU_energy_per_perticle_SPO1}
\frac{E}{N+1}=\frac{1}{4}\mathcal{C}^{2}(2+\mathcal{C}^{2}\beta)=\frac{\kappa^{2}(1-\kappa^{2})}{\beta(1-2{\kappa}^{2})^{2}}
\end{equation}
by substituting simply the solution $\hat{x}(t)$ of Eq. (\ref{sol_FPU_single_equation_SPO1}) in Hamiltonian
(\ref{FPU_Hamiltonian_beta}).

SPO2 is defined in a similar way. In particular, it exists for $N=5+3m,\;m=0,1,2,\ldots$ and corresponds to the case where every third particle is fixed, while the two in between move in opposite directions (in an out of phase fashion). Following similar arguments as for the SPO1 mode, the energy per particle of SPO2 is given by \cite{Antonopoulosetal2006a}:
\begin{equation}\nonumber
\frac{E}{N+1}=\frac{2\kappa^{2}(1-{\kappa}^{2}) }{3\beta(1-2{\kappa}^{2})^{2}}.
\end{equation}
We treat $E$ as a control parameter for the chaoticity of the FPU system \eqref{FPU_Hamiltonian_beta}. From now on, we drop the time-dependence notation of all involved variables for simplicity but use it wherever is needed.

\subsection{Bose-Einstein condensate Hamiltonian}\label{BEC_section}

The second Hamiltonian system we use in this paper is the Bose-Einstein Condensate (BEC) model \cite{Antonopoulosetal2006b} which is given by:
\begin{equation}\label{BEC_Hamiltonian_2}
H=\frac{1}{2}\sum_{j=1}^{N}(\dot{x}_{j}^{2}+x_{j}^{2})+\frac{1}{8}\sum_{j=1}^{N}(\dot{x}_{j}^{2}+x_{j}^{2})^2-\frac{1}{2}\sum_{j=1}^{N}(\dot{x}_{j}\dot{x}_{j+1}+x_{j}x_{j+1}),
\end{equation}
where $x_{j}$, $\dot{x}_{j}$ is the position and conjugate momentum of the $j$th particle (i.e. boson), respectively.

It possesses the second integral of motion:
\begin{equation}\label{BEC_Hamiltonian_second_integral}
F=\sum_{j=1}^{N}(\dot{x}_{j}^{2}+x_{j}^{2}),
\end{equation}
and therefore chaotic behavior can only occur for $N\geq3$.

We impose periodic boundary conditions in Eq.
(\ref{BEC_Hamiltonian_2}):
\begin{eqnarray}\label{BEC_periodic_boundary_conditions}
x_{N+1}(t)&=&x_{1}(t)\;\mathrm{and}\nonumber\\
\dot{x}_{N+1}(t)&=&\dot{x}_{1}(t),\;\forall t,
\end{eqnarray}
and use, for the same reason as in the FPU case, initial conditions set in the neighborhood of the out-of-phase mode (OPM):
\begin{eqnarray}\label{BEC_non_lin_mode_periodic_boundary_conditions_OPM}
x_{j}(t)&=&-x_{j+1}(t)\equiv\hat{x}(t),\nonumber\\
\dot{x}_{j}(t)&=&-\dot{x}_{j+1}(t)\equiv\dot{\hat{x}}(t),\;\forall
j=1,\ldots,N
\end{eqnarray}
with $N$ being even.

\subsection{Observation subspaces and quantities calculated on them}\label{Observational_Subspaces_section}

The FPU system \eqref{FPU_Hamiltonian_beta} can be simply written in the form:
\begin{equation}\label{KplusP_equation}
H=K+P=E=\mbox{const}
\end{equation}
where:
\begin{eqnarray}
K&=&\frac{1}{2}\sum_{j=1}^{N}\dot{x}_{j}^{2}\mbox{ and}\nonumber\\
P&=&\sum_{j=1}^{N}\biggl
(\frac{1}{2}(x_{j+1}-x_{j})^2+\frac{1}{4}\beta(x_{j+1}-x_{j})^4\biggr).\label{potential_nrg}
\end{eqnarray}
However, the BEC system \eqref{BEC_Hamiltonian_2} is not written in the same form and this will allow us to generalize the results of our study in the case where the $KP$ space is not implied directly by the Hamiltonian form.

In our analysis, we define and study quantities like Lyapunov exponents initially in the bi-dimensional $KP$ space, since $K$ is a meaningful physical quantity. Potential energy can be easily measured as well or estimated since $P=E-K$. However, we also consider the $(x_1,x_N)$ observation space which is constructed by the position coordinates of the first and last particle of the Hamiltonian. For the FPU case, we know that:
\begin{equation}\label{Hamiltonian_time_derivative}
\frac{dH}{dt}=\frac{dK}{dt}+\frac{dP}{dt}=0,
\end{equation}
where:
\begin{eqnarray}
\frac{dK}{dt}&=&\sum_{j=1}^N\dot{x_j}\ddot{x_j}\mbox{ and}\label{Kinetic_time_derivative}\\
\frac{dP}{dt}&=&\sum_{j=1}^N\Bigr[(\dot{x}_{j+1}-\dot{x}_j)\bigr[(x_{j+1}-x_j)+\beta(x_{j+1}-x_j)^3\bigl]\Bigr].\nonumber
\end{eqnarray}
Equation \eqref{Hamiltonian_time_derivative} is valid since the FPU Hamiltonian \eqref{FPU_Hamiltonian_beta} is a global integral of the motion and thus a conserved quantity during time evolution.

Along the lines of ideas presented in Ref. \cite{Baptistaetal2012}, we compute the upper bound $I_c$ for the $\mbox{MIR}$ between any two groups of $N/2$ nodes each. The upper bound $I_c$ for the $\mbox{MIR}$ is defined as (see supplementary material in Ref. \cite{Baptistaetal2012}):
\begin{equation}\label{IcHamiltonian}
I_c=2\sum_{i=1}^{\tilde{N}}\tilde{\lambda}_i-\tilde{H}_{\mbox{KS}}=2\tilde{H}-\tilde{H}_{\mbox{KS}},
\end{equation}
where $\tilde{N}$ is half the number of positive Lyapunov exponents measured in the subspace. Naturally, $\tilde{N}\leq N/2$. However, for the simulations we have performed we have set $\tilde{N}=N/2$. So, $\tilde{\lambda}_i,\;i=1,\ldots,N$ represent the greater than or equal to zero Lyapunov exponents of the $N$-dimensional projection constructed using scalar time series $x_i$, for $i=1,\ldots,N$, which can be calculated in many ways, for example by calculating the finite size and finite time Lyapunov exponents or expansion rates \cite{Baptistaetal2012}. $\tilde{H}_{\mbox{KS}}=\sum_{i=1}^{N}\tilde{\lambda}_i$ represents the sum of all greater than or equal to zero Lyapunov exponents of the projection (i.e. an approximation for the KS entropy) and $\tilde{H}=\sum_{i=1}^{\tilde{N}}\tilde{\lambda}_i$. Herein, we estimate them by computing the Lyapunov exponents of the Hamiltonian following \cite{Benettin1980a,Benettin1980b} and by keeping only those that are positive.

We also need to compute the upper bound $I_c^{KP}$ for the MIR in the bi-dimensional $KP$ space representing the maximum information exchanged between the kinetic ($K$) and potential ($P$) energies. Using the ideas from Ref. \cite{Baptistaetal2012}, $I_c^{KP}$ is given by:
\begin{equation}\label{I_cKP}
 I_c^{KP}=\lambda_1^{KP}-\lambda_2^{KP}
\end{equation}
where $\lambda_1^{KP}$ and $\lambda_2^{KP}$ are the two positive Lyapunov exponents of the $KP$ space with $\lambda_1^{KP}>\lambda_2^{KP}$.  In the case where $\lambda_2^{KP}\leq0$, we have $I_c^{KP}=\lambda_1^{KP}$ and thus it turns out that $\mbox{MIR}^{KP}\leq\lambda_1^{KP}$ (see Ref. \cite{Baptistaetal2012}).

In a series of papers \cite{Rechesteretal1979,Pettinietal1990,Pettinietal1991,Pettinietal2005,Benettin1984,Livietal1986,Antonopoulosetal2006b}, the authors report for dynamical systems ranging from different kinds of billiards to multi-dimensional Hamiltonian systems, that the largest Lyapunov exponent $\lambda_1$ of the system scales with the energy $E$ with a power-law of the form:
\begin{equation}\label{LE1hampowerlawE}
 \lambda_1\propto E^{b}
\end{equation}
where $b$ is a real positive constant. This power-law dependence is valid for a rather large energy interval that can support chaotic behavior.

To numerically calculate $\frac{dK}{dt}$ we use:
\begin{equation}\nonumber
\frac{dK}{dt}\approx\frac{K(t)-K(t-dt)}{dt}=\frac{\Delta K}{dt} 
\end{equation}
from which we can define the time average of the absolute value of the transfer of kinetic energy per unit of time through:
\begin{equation}\label{averagetimeKdot}
	\langle\biggl|\frac{dK}{dt}\biggr|\rangle_t\approx\langle\biggl|\frac{\Delta K}{dt}\biggr|\rangle_t,
\end{equation}
where $\langle\cdot\rangle_t$ denotes the time average over the integration of the trajectory $\vec{X}(t)$ up to $t=t_{\mbox{f}}$. $|\cdot|$ is the absolute value of the argument and we use it because we want to relate the quantities of Eq. \eqref{averagetimeKdot} to positive average quantities, such as the positive Lyapunov exponents. Accordingly, $\Delta K$ is the amount of kinetic energy being transferred between $K$ and $P$ during a time step.

Since the BEC Hamiltonian \eqref{BEC_Hamiltonian_2} is not of the form $H=K+P$ as the FPU system, we reside on the calculation of a similar quantity $\langle\bigl|\frac{\Delta K_1}{dt}\bigr|\rangle_t$ based on the kinetic energy of any of its particles, for example of the first particle $x_1$:
\begin{equation}\label{Delta_K1}
\langle\biggl|\frac{\Delta K_1}{dt}\biggr|\rangle_t=\langle\biggl|\frac{K_1(t)-K_1(t-dt)}{dt}\biggr|\rangle_t,
\end{equation}
where $K_1=\frac{1}{2}\dot{x}_{1}^{2}$ is the kinetic energy of the first particle. Equation \eqref{Delta_K1} is similar to the quantity $\langle\bigl|\frac{\Delta K}{dt}\bigr|\rangle_t$ of the left hand side of Eq. \eqref{DeltaKpowerlawE}.

\subsection{Set of initial conditions}\label{ic_section}

We prepare the two systems in a systematic way to reside in a chaotic regime and be able to produce information. For example, for the SPO2 we follow Ref. \cite{Antonopoulosetal2006a} and consider $\beta=1$ and $N=14$ varying the energy and initial condition $\vec{X}(0)$ appropriately as following: For each fixed energy $E$ of Hamiltonian \eqref{FPU_Hamiltonian_beta}, an initial condition $\vec{X}(0)=(\vec{x}(0),\dot{\vec{x}}(0))$ is chosen (where $\vec{x}(t)=(x_1(t),x_2(t),\ldots,x_N(t))$ and $\dot{\vec{x}}(t)=(\dot{x_1}(t),\dot{x_2}(t),\ldots,\dot{x_N}(t))$) so that it lies in the neighborhood of SPO2. By neighborhood we mean that we perturb the equations of motion by a controllable small perturbation (i.e. $\hat{x}(t)=\hat{x}(t)-10^{-15}$) so that the perturbed initial condition $\vec{X}(0)$ will be at the same constant energy $E$ of SPO2.  Easily, we can fullfil this requirement by solving Eq. \eqref{FPU_Hamiltonian_beta} for $\dot{x}(N)$ and then substitute it in the initial condition. A demonstration of the importance of this can be found in Sec. \ref{results_section} where we present the relation between the largest Lyapunov exponent of the $KP$ space and of the FPU Hamiltonian.

We thus end up with 14 nodes, each interacting with its nearest neighbours in a 1-dimensional lattice with fixed ends. In our example, SPO2 is destabilized at the energy $E_{\mbox{u}}\approx0.117$ and restabilized again at $E_{\mbox{r}}\approx47.059$ \cite{Antonopoulosetal2006a}. Thus, as $E$ increases in $(E_{\mbox{u}},E_{\mbox{r}})$, SPO2 is unstable and gives rise initially to weakly and then to strongly chaotic behaviour in its neighborhood. For each $E$ we numerically integrate the corresponding initial condition $\vec{X}(0)$ and compute the Lyapunov exponents following Refs. \cite{Benettin1980a,Benettin1980b,Skokos2010} until they show a clear tendency to converge to a value. We subsequently record their values at the final integration time $t_{\mbox{f}}$. In our case, we have checked that this  convergence happens at about $t_{\mbox{f}}=2\times10^6$. We denote them as $\lambda_i,\;i=1,\ldots,N$ arranged in descending order. In terms of the numerical integration, we try to satisfy the condition that the relative energy error is kept between $10^{-6}$ and $10^{-13}$. We follow a similar approach for the initial conditions we set in the neighborhood of SPO1 mode of FPU and OPM mode of BEC so that we can guarantee chaotic behavior with the increase of the energy of the system.

\section{Results}\label{results_section}

\subsection{Relation between largest Lyapunov exponent of the bi-dimensional $KP$ space and of the Hamiltonian}\label{section_Results_2013_02_12_A}

The dynamics on the $KP$ space is driven by the dynamics of the Hamiltonian system and we have no explicitly given equations of motion for the $KP$ space. As we have already pointed out, we choose initial conditions $\vec{X}(0)$ on the same energy as the SPO2, and this implies that points $(K(\vec{X_1}(t)),P(\vec{X_1}(t)))$ and $(K(\vec{X_2}(t)),P(\vec{X_2}(t)))$ belong to the line $K(\vec{X}(t))+P(\vec{X}(t))=E$. The motion takes place on this 1-dimensional subspace and thus, there is only one Lyapunov exponent $\lambda_1^{KP}$ that leads to $I_c^{KP}=\lambda_1^{KP}$.

\begin{figure}[!ht]
\begin{flushleft}
\includegraphics[width=16.15cm,height=10cm]{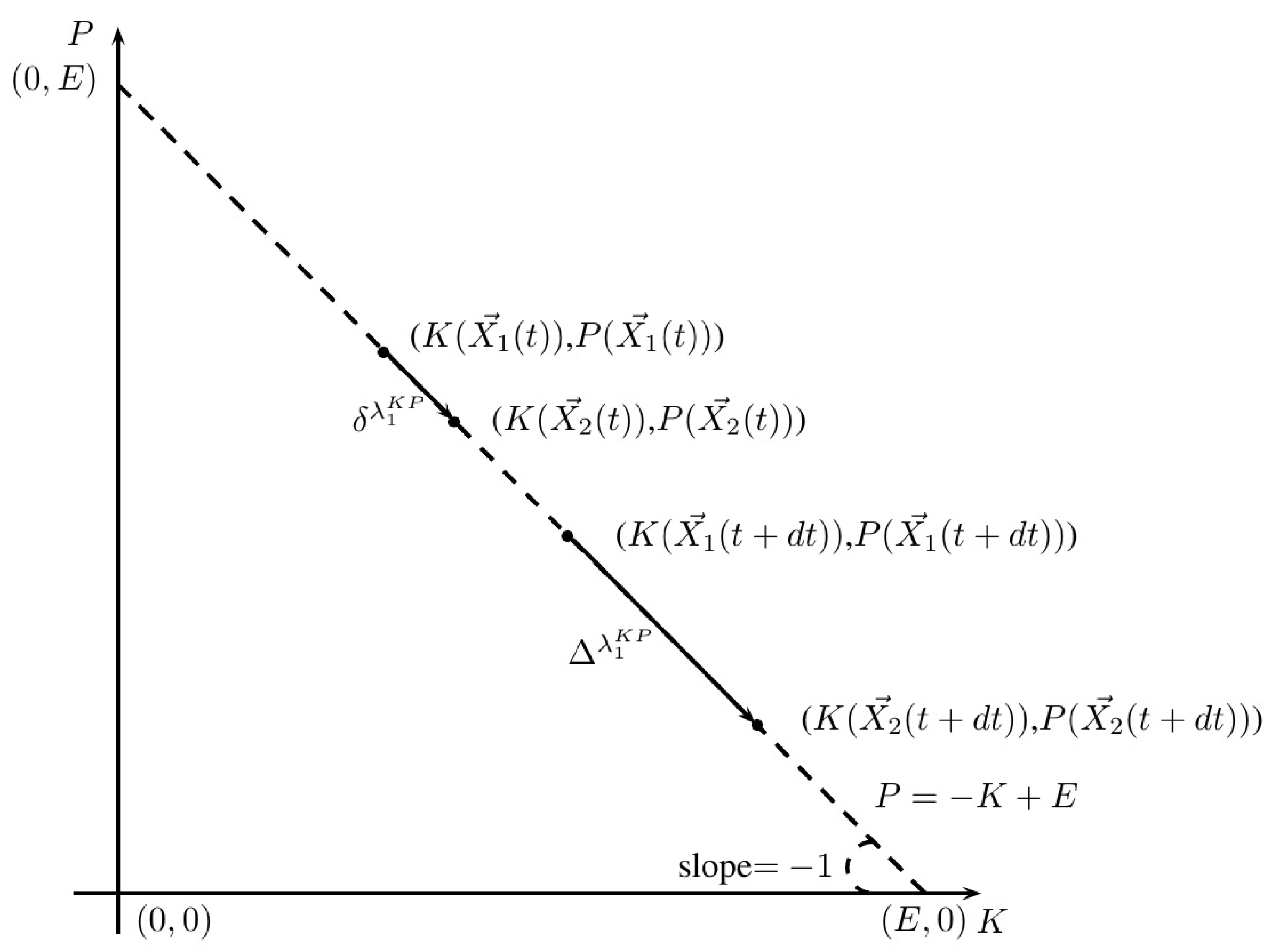}
\end{flushleft}
\caption{Schematic representation of the time evolution after one time step $dt$ of two deviation vectors (arrows) corresponding to the direction along the Lyapunov exponent $\lambda_1^{KP}$ on the 1-dimensional subspace $K+P=E$ on the $KP$ space. $\vec{X_1}(t)$ and $\vec{X_2}(t)$ are two trajectories in the phase space of Hamiltonian \eqref{FPU_Hamiltonian_beta} that drive the dynamics along this line. We denote with $\delta$ and $\Delta$ the lengths of the two deviation vectors initially and after one time step, respectively.}\label{LEs_KP_plane}
\end{figure}

In Fig. \ref{LEs_KP_plane}, one can see schematically the time evolution after one time step $dt$ of a deviation vector (denoted as an arrow) along the direction of the Lyapunov exponent $\lambda_1^{KP}$ defined for the dynamics on the line $K+P=E$. Here $\vec{X_1}(t)$ and $\vec{X_2}(t)$ are two trajectories in the phase space of Hamiltonian \eqref{FPU_Hamiltonian_beta} on the same energy $E$ as SPO2, started initially in its neighborhood and being infinitesimally close. Then, $\lambda_1^{KP}$ is the rate of expansion of the deviation vector defined by the points $(K(\vec{X_1}(t))$,$P(\vec{X_1}(t)))$ and $(K(\vec{X_2}(t))$,$P(\vec{X_2}(t)))$. Here, $\delta$ and $\Delta$ denote the lengths of the initial and after one time step deviation vectors respectively.

$\lambda_1^{KP}$ can be defined for infinitesimally close-by points on the 1-dimensional space of $K+P=E$ of Fig. \ref{LEs_KP_plane} by keeping track of the evolution of their distance. In particular, for such points $(K(\vec{X_1}(t))$,$P(\vec{X_1}(t)))$ and $(K(\vec{X_2}(t))$,$P(\vec{X_2}(t)))$, their distance is given by:
\begin{eqnarray}
	\Delta K_X(t)^2=(K(\vec{X_1}(t))-K(\vec{X_2}(t)))^2+(P(\vec{X_1}(t))-P(\vec{X_2}(t)))^2\nonumber=\\
	(K(\vec{X_1}(t))-K(\vec{X_2}(t)))^2+(E-K(\vec{X_1}(t))-E+K(\vec{X_2}(t)))^2\nonumber=\\
	2(K(\vec{X_1}(t))-K(\vec{X_2}(t)))^2\nonumber\Rightarrow\\
	\Delta K_X(t)=\sqrt{2}|K(\vec{X_1}(t))-K(\vec{X_2}(t))|.\label{DeltaK_LE1_KP_plane}
\end{eqnarray}
Defining:
\begin{equation}\nonumber
	\lambda_1^{KP}=\lim_{t\rightarrow\infty}\frac{1}{t}\log\Biggl({\frac{\Delta K(t)}{\Delta K(0)}}\Biggr)\mbox{ for }\Delta K(0)\rightarrow0,
\end{equation}
and combining it with Eq. \eqref{DeltaK_LE1_KP_plane} we obtain:
\begin{equation}\label{lambda1KP_2_2}
	\lambda_1^{KP}=\lim_{t\rightarrow\infty}\frac{1}{t}\log\Biggr(\frac{|K(\vec{X_1}(t))-K(\vec{X_2}(t))|}{|K(\vec{X_1}(0))-K(\vec{X_2}(0))|}\Biggl).
\end{equation}

\begin{figure}[!ht]
\begin{flushleft}
\includegraphics[width=16.15cm,height=10cm]{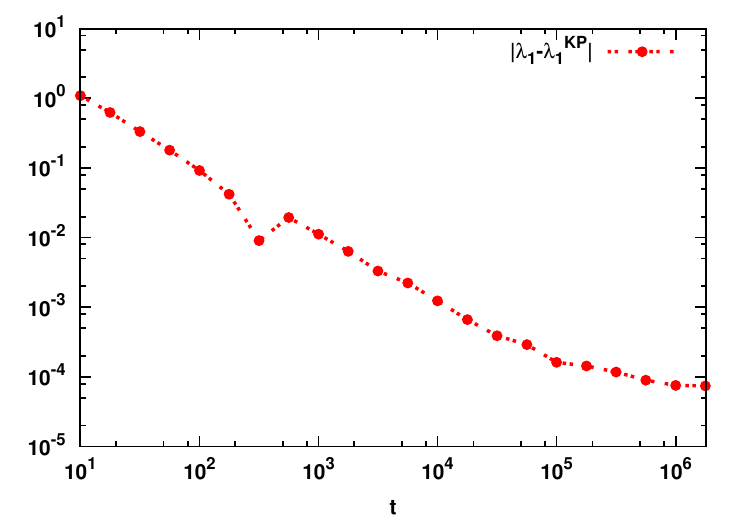}
\end{flushleft}
\caption{Plot of the absolute difference $|\lambda_1-\lambda_1^{KP}|$ as a function of time for two trajectories $\vec{X_1}(t)$ and $\vec{X_2}(t)$ located initially in the neighborhood of SPO2 at the same energy $E$. Here, $E=30$ is well inside the interval $(E_{\mbox{u}},E_{\mbox{r}})$. Note that both axes are logarithmic.}\label{absLE_1KPminusLE_1H}
\end{figure}

We denote as $\lambda_1$ the largest Lyapunov exponent in the neighborhood of SPO2, and reside on numerical simulations to show in Fig. \ref{absLE_1KPminusLE_1H} that $\lambda_1^{KP}$ is actually $\lambda_1$. In the example of Fig. \ref{absLE_1KPminusLE_1H} we have set $E=30$, resulting in the relation $K+P=30$. However, we have checked that the above result is valid for all energies we considered in $(E_{\mbox{u}},E_{\mbox{r}})$. We observe that $|\lambda_1-\lambda_1^{KP}|$ tends to zero in the course of time and that at some point starts to saturate at about $10^{-4}$ due to round off numerical errors. In other words, we have showed that the largest Lyapunov exponent of the 1-dimensional $K+P=E$ space is equal to the largest Lyapunov exponent $\lambda_1$ of Hamiltonian \eqref{FPU_Hamiltonian_beta}, i.e. $\lambda_1^{KP}=\lambda_1$.

To achieve this result, we integrated simultaneously two infinitesimally close trajectories $\vec{X_1}(t)$ and $\vec{X_2}(t)$ (e.g. at an initial distance of the order of $10^{-7}$) on the same energy as SPO2 and consider thus that $\Delta K(0)\approx10^{-7}$, and replace the limits in Eq. \eqref{lambda1KP_2_2} by a finite time $t=2\times10^6$, computing $\lambda_1^{KP}$ as a time average \cite{Benettin1980b}, i.e. as finite size and finite time Lyapunov exponent. Since for chaotic trajectories, the distance between $\vec{X_1}(t)$ and $\vec{X_2}(t)$ quickly saturates, we periodically renormalize their separation without altering their relative orientation in phase space and then compute the new distance $|K(\vec{X_1}(t))-K(\vec{X_2}(t))|$ setting $|K(\vec{X_1}(0))-K(\vec{X_2}(0))|=|K(\vec{X_1}(t-dt))-K(\vec{X_2}(t-dt))|$. To avoid any numerical overflows, we preferred to do this at every time step.

Note that $\lambda_1^{KP}=\lambda_1$ is not an unexpected result, since the largest Lyapunov exponent should be obtained in typical low-dimensional linear projections or embedding spaces \cite{Eckmann1985,WOL85}. By typical here we mean bi-dimensional subspaces or projections that are not oriented along Lyapunov vectors. However, the $KP$ space is a highly nonlinear projection still maintaining the largest positive Lyapunov exponent of the Hamiltonian as we have demonstrated. Every initial condition creates a trajectory with only one positive Lyapunov exponent in the $KP$ subspace. Therefore, $I_c^{KP}=\lambda_1^{KP}$.

Concluding this part, we have demonstrated that the transfer of information from $K$ to $P$ is mediated by the largest Lyapunov exponent of the Hamiltonian. We finally obtain:
\begin{equation}\nonumber
\mbox{MIR}^{KP}\leq I_c^{KP}=\lambda_1^{KP}=\lambda_1.
\end{equation}
The last result implies that the upper bound $I_c^{KP}$ for the MIR$^{KP}$ between kinetic and potential energies is equal to the largest Lyapunov exponent of the Hamiltonian and consequently, $\mbox{MIR}^{KP}$ can not be bigger than this exponent.

\subsection{Relation between production and transfer of information in the small energy regime}\label{section_Results_2013_02_12_B}

\begin{figure}[!ht]
\begin{flushleft}
\includegraphics[width=16.15cm,height=10cm]{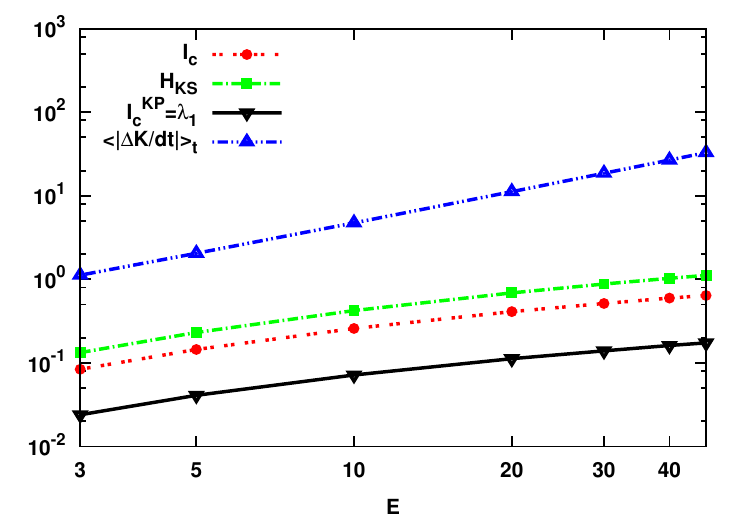}
\end{flushleft}
\caption{Plot of the quantities: $I_c$ as defined by Eq. \eqref{IcHamiltonian} in red dashed line with points, $H_{\mbox{KS}}$ as defined by Eq. \eqref{HKS_entropy} in green dashed line with rectangles, $I_{c}^{KP}$ as defined by Eq. \eqref{I_cKP} in black solid line with lower triangles and $\langle|\frac{\Delta K}{dt}|\rangle_t$ as defined by Eq. \eqref{averagetimeKdot} in blue dashed line with upper triangles as a function of $E$ for initial conditions $\vec{X}(0)$ located in the neighborhood of SPO2 of the FPU system. Note that both axes are logarithmic.}\label{IcLE_and_maxdK_dt_vs_nrg_Ham}
\end{figure}

To start with, we present in a log-log plot in Fig. \ref{IcLE_and_maxdK_dt_vs_nrg_Ham} the quantities $I_c$ of Eq. \eqref{IcHamiltonian} in red dashed line with points, $H_{\mbox{KS}}$ of Eq. \eqref{HKS_entropy} in green dashed line with rectangles, $I_c^{KP}=\lambda_1$ in black solid line with lower triangles and $\langle|\frac{\Delta K}{dt}|\rangle_t$ of Eq. \eqref{averagetimeKdot} in blue dashed line with upper triangles for the SPO2 case of the FPU system with parameters as defined in Subsec. \ref{FPU_section}. Here $dt$ is the time step of the integration (i.e. $dt\ll1$). The time derivative of the kinetic energy $\frac{dK}{dt}$ accounts for the rate of transfer from kinetic to potential energy. We see that all quantities follow the same morphology (i.e. share the same functional form) as the energy of the initial condition $\vec{X}(0)$ is increased in the interval $(E_{\mbox{u}},E_{\mbox{r}})$. Moreover, $H_{\mbox{KS}}$ is an upper bound of the upper bound $I_c$ for the $\mbox{MIR}$ between two groups formed each by 7 nodes. We will prove a related inequality in the Appendix.

The approach we shall follow to relate $\langle|\frac{\Delta K}{dt}|\rangle_t$ with $I_c^{KP}$ for the transfer of information between $K$ and $P$ is meaningful as long as the motion in the Hamiltonian phase space is chaotic (e.g. as long as $E\in(E_{\mbox{u}},E_{\mbox{r}})$). If the motion is periodic or quasi-periodic there is no exchange of information between the nodes (i.e. by knowing the position of a particular node one can predict the position and momenta of another one). Our results show that $\langle|\frac{\Delta K}{dt}|\rangle_t$ is related by a power-law to the largest Lyapunov exponent $\lambda_1$ of the Hamiltonian and to the upper bound $I_c^{KP}$ for the transfer of information between kinetic and potential energies. Surprisingly, we have found that this is valid for sufficiently large enough subintervals, i.e. for $E\in(E_{\mbox{u}},E_{\mbox{r}})$.

Here, we need to make use of only one neighboring initial condition $\vec{X}(0)$ of SPO2 and denote for simplicity by $K(t)\equiv K(\vec{X}(t))$. With the help of Eq. \eqref{averagetimeKdot} and $\Delta K\equiv\Delta K(t)=K(t)-K(t-dt)$ we have found numerically that:
\begin{equation}\label{DeltaKpowerlawE}
	\langle\biggl|\frac{\Delta K}{dt}\biggr|\rangle_t\propto E^{b_2}
\end{equation}
for the same energy interval that Eq. \eqref{LE1hampowerlawE} applies where $b_2$ is a real positive constant. By substituting Eq. \eqref{LE1hampowerlawE} in Eq. \eqref{DeltaKpowerlawE}, we obtain:
\begin{equation}\label{first_result}
	\langle\biggl|\frac{\Delta K}{dt}\biggr|\rangle_t\propto\Bigr({I_c^{KP}}\Bigl)^{\frac{b_2}{b_1}}=\biggl({\lambda_1^{KP}}\biggr)^{\frac{b_2}{b_1}},
\end{equation}
where we have used $I_c^{KP}=\lambda_1$ (see Subsec. \ref{section_Results_2013_02_12_A}). It is straightforward to show that the same power-law \eqref{first_result} applies to $\langle\bigl|\frac{\Delta P}{dt}\bigr|\rangle_t$ due to Eqs. \eqref{KplusP_equation} and \eqref{Hamiltonian_time_derivative} respectively. We emphasize that $\bigl|\frac{\Delta K}{dt}\bigr|$ is a time-ratio that depends on time, and that $\langle\bigl|\frac{\Delta K}{dt}\bigr|\rangle_t$ and ${\lambda_1}$ are time invariant averages.

\begin{figure}[!ht]
\begin{flushleft}
\includegraphics[scale=1.2]{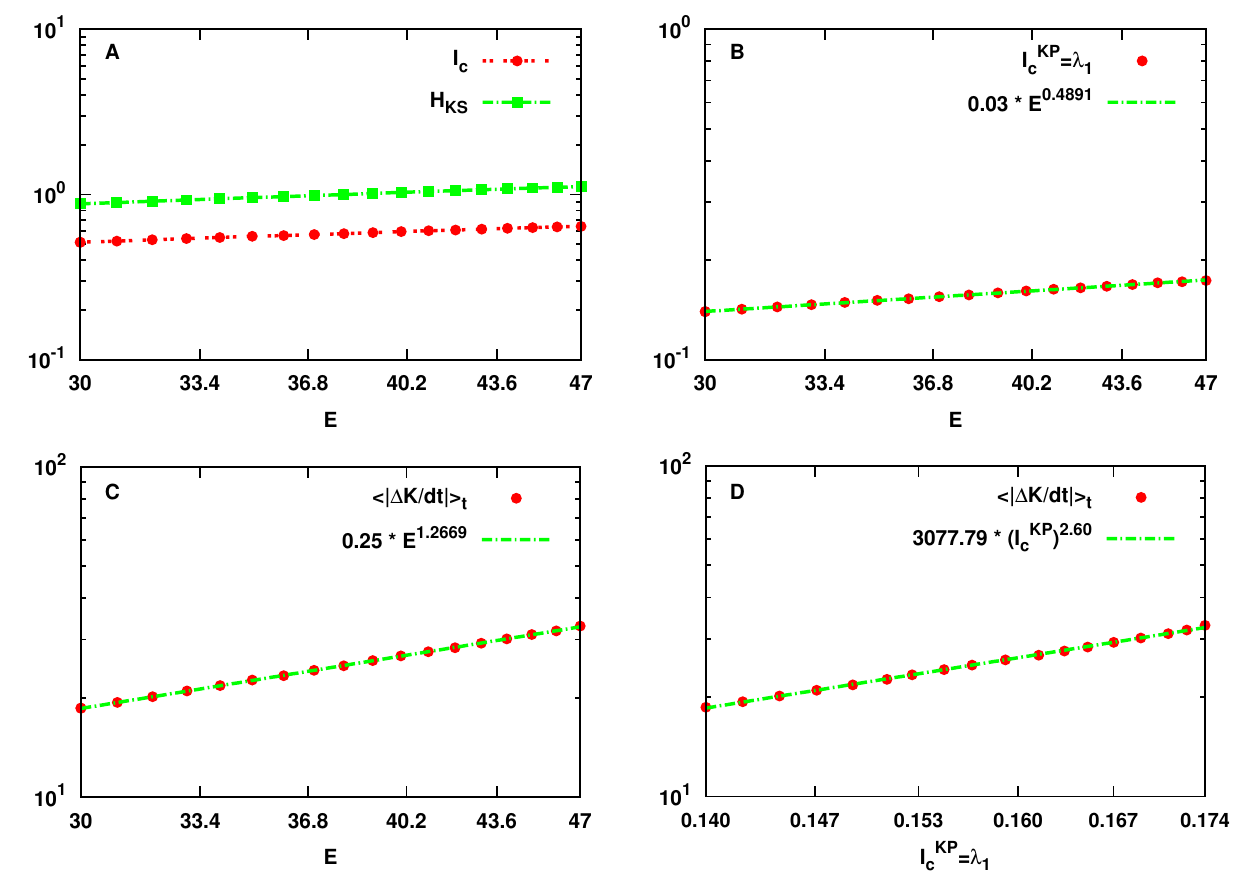}
\end{flushleft}
\caption{Panel A: Plot of quantities: $I_c$ of Eq. \eqref{IcHamiltonian} in red dashed line with points and $H_{\mbox{KS}}$ of Eq. \eqref{HKS_entropy} in green dashed line with rectangles. Panel B: Plot of quantities $I_c^{KP}=\lambda_1$ with red points with the power-law fitting of Eq. \eqref{LE1hamEfit} in green line. Panel C: Plot of $\langle\bigl|\frac{\Delta K}{dt}\bigr|\rangle_t$ with red points with the power-law fitting of Eq. \eqref{timeaverageDKLE1fit} in green line. Panel D: Power-law dependence of $\langle\bigl|\frac{\Delta K}{dt}\bigr|\rangle_t$ to $I_c^{KP}=\lambda_1$ in red points, in the interval $(0.140,0.174)$ that corresponds to the energy interval $[30,47]$ of panels A, B and C and of the power-law fitting of Eq. \eqref{timeaverageDKLE1fit} in green dashed line. Note that all axes are logarithmic.}\label{IcLE_and_maxdK_dt_vs_nrg_COMBINED_ndof=14_1}
\end{figure}

Fig. \ref{IcLE_and_maxdK_dt_vs_nrg_COMBINED_ndof=14_1}A shows in a log-log scale the quantity $I_c$ of Eq. \eqref{IcHamiltonian} in red dashed line with points and $H_{\mbox{KS}}$ of Eq. \eqref{HKS_entropy} in green dashed line with rectangles. In panel B, we plot $I_c^{KP}=\lambda_1$ with red points and the power-law fitting:
\begin{equation}
	\lambda_1=a_1 E^{b_1}\label{LE1hamEfit}
\end{equation}
with green line. The agreement is remarkable. In Fig. \ref{IcLE_and_maxdK_dt_vs_nrg_COMBINED_ndof=14_1}C we plot $\langle\bigl|\frac{\Delta K}{dt}\bigr|\rangle_t$ of Eq. \eqref{DeltaKpowerlawE} and fit with the power-law:
\begin{equation}
	\langle\biggl|\frac{\Delta K}{dt}\biggr|\rangle_t=a_2 E^{b_2}\label{DKEfit}
\end{equation}
showed as green line. We find that $a_1\approx0.03,\;b_1\approx0.489$ and that $a_2\approx0.25,\;b_2\approx1.267$. In panel D of the same figure we plot $\langle\bigl|\frac{\Delta K}{dt}\bigr|\rangle_t$ with red points as a function of $I_c^{KP}=\lambda_1$ for values that correspond to the same energy interval of panels A, B and C. The power-law fitting:
\begin{equation}
	\langle\biggl|\frac{\Delta K}{dt}\biggr|\rangle_t\approx a_3\bigl(\lambda_1\bigr)^{b_3}\label{timeaverageDKLE1fit}
\end{equation}
plotted in green dashed line gives $a_3\approx3077.79$ and $b_3\approx2.60$ which is in good agreement with the value of $b_2/b_1\approx2.591$. The above arguments directly imply that:
\begin{equation}\label{eq:main_result_1}
	\langle\biggl|\frac{\Delta K}{dt}\biggr|\rangle_t\propto\Bigr({I_c^{KP}}\Bigl)^{\frac{b_2}{b_1}}=\Biggr(\frac{2H_{\mbox{KS}}}{N}\Biggl)^{\frac{b_2}{b_1}},
\end{equation}
where the proportionality constant $a_3=a_2\bigl(\frac{1}{a_1}\bigl)^{\frac{b_2}{b_1}}$ and $b_3=b_2/b_1$. To arrive at Eq. \eqref{eq:main_result_1} we have used $I_c^{KP}=\lambda_1$ of Subsec. \ref{section_Results_2013_02_12_A} and Eq. \eqref{HKSintegral} presented in the Appendix. Equation \eqref{eq:main_result_1} relates the production $\lambda_1$ and transfer of information $I_c^{KP}$ in the $KP$ space with $\langle\bigl|\frac{\Delta K}{dt}\bigr|\rangle_t$ and $H_{\mbox{KS}}$. Therefore, the larger the transfer of energy is between the kinetic and potential energy, the larger is the upper bound for the MIR between the kinetic and potential energies and the larger the KS entropy of the system will be. In other words, exchange of information between $K$ and $P$ implies exchange of energy, and vice-versa. However, a relatively small increment of energy transfer produces a larger relative increase of the information transferred since $b_3>1$.

In the Appendix we prove another important result which is the inequality:
\begin{equation}
I_c^{KP}<I_c<H_{\mbox{KS}}, 
\end{equation}
and thus justify the result presented in Fig. \ref{IcLE_and_maxdK_dt_vs_nrg_Ham}.

\subsection{Generalization of our study}\label{large_nrg_regime_sec}

Here, we  extend our study and present the generalization of our predicted upper bounds for the MIR and the connection with the transfer of energy of the previous section by considering higher energy intervals with initial conditions set in different parts of the phase space of two Hamiltonian systems: the FPU \eqref{FPU_Hamiltonian_beta} and BEC \eqref{BEC_Hamiltonian_2}.

We will show that if one considers a much larger energy interval for these systems with initial conditions set in different parts of their phase spaces, then Eqs. \eqref{LE1hamEfit}, \eqref{DKEfit} and \eqref{timeaverageDKLE1fit} can be generalized, as:
\begin{eqnarray}
 I_c^{\mbox{BS}}&=&a_4+b_4E^{c_4},\;\;a_4,b_4,c_4\in\mathbb{R},\label{LE1hamEfit_big_energy}\\
 \langle\biggl|\frac{\Delta K}{dt}\biggr|\rangle_t&=&a_5+b_5(c_5+E)^{d_5},\;\;a_5,b_5,c_5,d_5\in\mathbb{R}\label{timeaverageDKLE1fit_big_energy}.
\end{eqnarray}
We prefer to call Eqs. \eqref{LE1hamEfit_big_energy} and \eqref{timeaverageDKLE1fit_big_energy} as generalized power-law functions. Here, $\mbox{BS}$ stands for the bi-dimensional space of observation. In the case of the FPU system \eqref{FPU_Hamiltonian_beta} we consider as a bi-dimensional space the $KP$ space while for the BEC system \eqref{BEC_Hamiltonian_2} we consider the observation space constructed by observing the pair of variables $x_1$ and $x_N$, that is by the position of the first and last particle. In Sec. \ref{Ham_com_chaneel_FPU}, where we study an ``experimental'' setup of a 1-dimensional communication channel based on the FPU system, we will use this particular observation space as well.

By eliminating $E$ from both Eqs. \eqref{LE1hamEfit_big_energy} and \eqref{timeaverageDKLE1fit_big_energy}, one arrives at the relation between transfer of energy per unit of time (i.e. $\langle\bigl|\frac{\Delta K}{dt}\bigr|\rangle_t$) and upper bound of information transmitted in the bi-dimensional space $\mbox{BS}$ (i.e. $I_c^{\mbox{BS}}$):
\begin{equation}
 f(I_c^{\mbox{BS}})=\langle\biggl|\frac{\Delta K}{dt}\biggr|\rangle_t=a_6+b_6\biggl[c_6+\biggl(\frac{d_6+I_c^{\mbox{BS}}}{e_6}\biggl)^{f_6}\biggr]^{g_6},\;\;a_6,b_6,c_6,d_6,e_6,f_6,g_6\in\mathbb{R}\label{eq:main_result_1_big_nrg_interval}.
\end{equation}
Parameters $a_i,b_i,c_i,d_i,e_i,f_i,g_i$ can be determined by performing a non-linear fitting of the numerical data by the functions \eqref{LE1hamEfit_big_energy}, \eqref{timeaverageDKLE1fit_big_energy} and \eqref{eq:main_result_1_big_nrg_interval}. We have used Matlab to perform these fittings.

\subsubsection{FPU SPO2}\label{large_nrg_regime_SPO2}

\begin{figure}[!ht]
\begin{flushleft}
\includegraphics[scale=1.2]{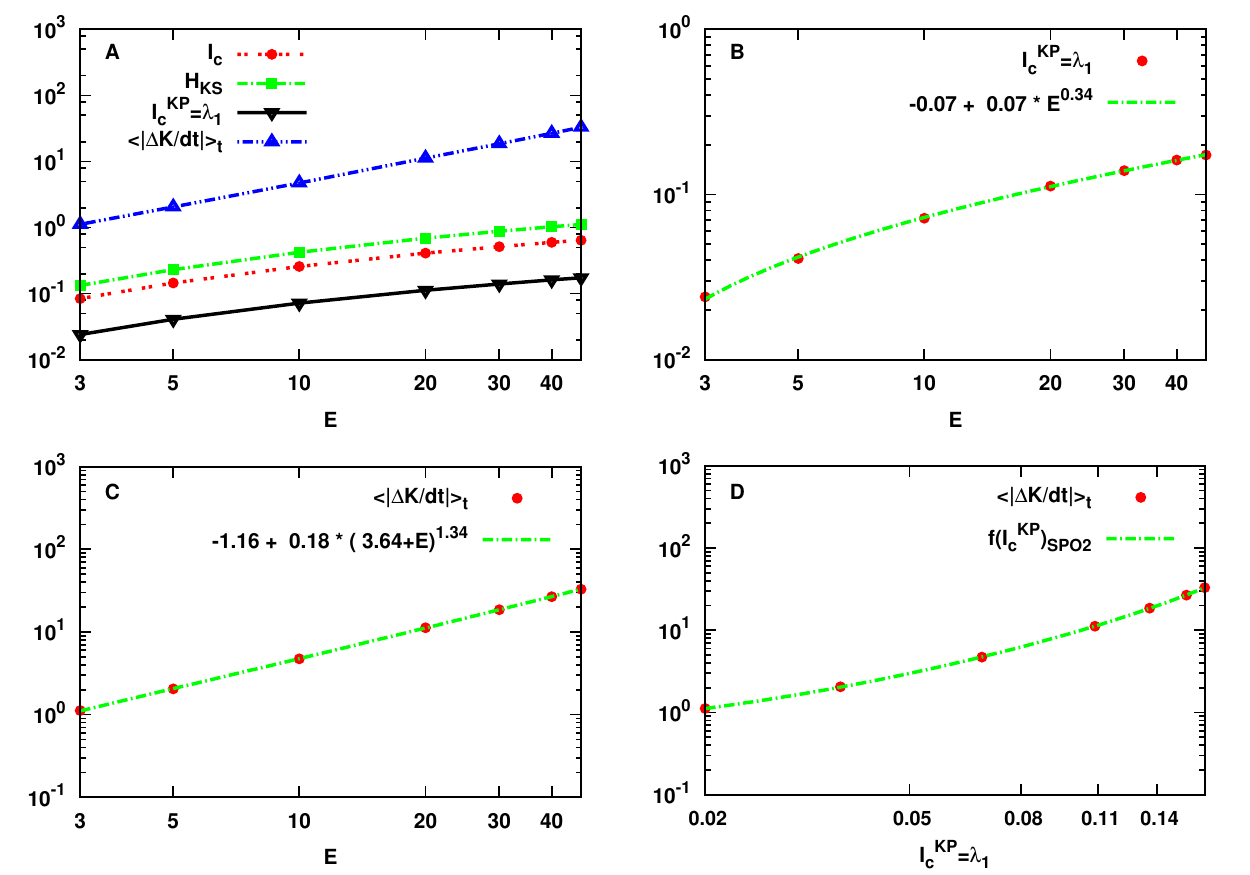}
\end{flushleft}
\caption{Panel A: Plot of quantities: $I_c$ as defined by Eq. \eqref{IcHamiltonian} in red dashed line with points, $H_{\mbox{KS}}$ as defined by Eq. \eqref{HKS_entropy} in green dashed line with rectangles, $I_{c}^{KP}$ as defined by Eq. \eqref{I_cKP} in black solid line with lower triangles and $\langle|\frac{\Delta K}{dt}|\rangle_t$ as defined by Eq. \eqref{averagetimeKdot} in blue dashed line with upper triangles as a function of $E$ for initial conditions $\vec{X}(0)$ located in the neighborhood of SPO2 of the FPU system. Note that both axes are logarithmic. Panel B: Plot of $I_c^{KP}=\lambda_1$ with red points with the power-law fitting of Eq. \eqref{LE1hamEfit_big_energy} in green line. Panel C: Plot of $\langle\bigl|\frac{\Delta K}{dt}\bigr|\rangle_t$ with red points with the power-law fitting of Eq. \eqref{timeaverageDKLE1fit_big_energy} in green line. Panel D: Power-law dependence of $\langle\bigl|\frac{\Delta K}{dt}\bigr|\rangle_t$ to $I_c^{KP}=\lambda_1$ in red points, in the interval $(0.02,0.174)$ that corresponds to the energy interval $[3,47]$ of panels A, B and C and of the power-law fitting of Eq. \eqref{timeaverageDKLE1fit_big_energy} in green dashed line. Note that all axes are logarithmic.
}
\label{IcLE_and_maxdK_dt_vs_nrg_COMBINED_ndof=14_big_nrg_interval_FPU_SPO2}
\end{figure}

In the case of the SPO2 studied in Subsec. \ref{section_Results_2013_02_12_B}, the fit of Fig. \ref{IcLE_and_maxdK_dt_vs_nrg_COMBINED_ndof=14_1} was performed in the energy interval $[30,47]$. Here we generalize Eqs. \eqref{LE1hamEfit}, \eqref{DKEfit} and \eqref{timeaverageDKLE1fit} in the larger energy interval $[3,47]$ for which the dynamics around SPO2 is chaotic as indicated by the Lyapunov exponents. This allows the creation and transfer of information and energy. We have used the same parameters and setup (e.g. 14 particles) to allow for a direct comparison between Figs. \ref{IcLE_and_maxdK_dt_vs_nrg_COMBINED_ndof=14_1} and \ref{IcLE_and_maxdK_dt_vs_nrg_COMBINED_ndof=14_big_nrg_interval_FPU_SPO2}.

By doing a similar analysis as in Subsec. \ref{section_Results_2013_02_12_B}, we present in Fig. \ref{IcLE_and_maxdK_dt_vs_nrg_COMBINED_ndof=14_big_nrg_interval_FPU_SPO2} the plots of all relevant quantities for the larger energy interval. By fitting the new data with the generalized power-laws of Eqs. \eqref{LE1hamEfit_big_energy}, \eqref{timeaverageDKLE1fit_big_energy} and \eqref{eq:main_result_1_big_nrg_interval} we have: $a_4\approx-0.07$, $b_4\approx0.07$, $c_4\approx0.34$ for Eq. \eqref{LE1hamEfit_big_energy}, $a_5\approx-1.16$, $b_5\approx0.18$, $c_5\approx3.64$, $d_5\approx1.34$ for Eq.\eqref{timeaverageDKLE1fit_big_energy} and finally: $a_6\approx-14.83$, $b_6\approx2.27\times10^{-15}$, $c_6\approx32.41$, $d_6\approx0.01$, $e_6\approx0.1$, $f_6\approx1.97$ and $g_6\approx10.47$ for Eq. \eqref{eq:main_result_1_big_nrg_interval}.

\subsubsection{FPU SPO1}\label{large_nrg_regime_SPO1}

Here we extend our study to a another part of the phase space of the FPU Hamiltonian with initial conditions set in the neighborhood of the periodic orbit SPO1 (see Eq. \eqref{FPU_non_lin_mode_fixed_boundary_conditions_SPO1} of Subsec. \ref{FPU_section}). We have chosen this particular part of the phase space as SPO1 does not restabilize at some bigger energy as it happens with SPO2 and thus allows to reach as high energies as desired. We will show that the same generalized power-laws of Eqs. \eqref{LE1hamEfit_big_energy}, \eqref{timeaverageDKLE1fit_big_energy} and \eqref{eq:main_result_1_big_nrg_interval} can still be used to fit the data of the upper bounds for MIR such as $I_c$, $H_{KS}$ and $I_c^{KP}$. In more details, for Eq. \eqref{LE1hamEfit_big_energy} we have: $a_4\approx-0.18$, $b_4\approx0.15$, $c_4\approx0.23$, for Eq.\eqref{timeaverageDKLE1fit_big_energy} we have: $a_5\approx-0.99$, $b_5\approx0.26$, $c_5\approx0$, $d_5\approx1.25$ and finally, for Eq. \eqref{eq:main_result_1_big_nrg_interval} we have: $a_6\approx-14.33$, $b_6\approx2.91$, $c_6\approx1.68$, $d_6\approx-0.08$, $e_6\approx0.17$, $f_6\approx1.39$ and $g_6\approx3.62$.

\begin{figure}[!ht]
\begin{flushleft}
\includegraphics[scale=1.2]{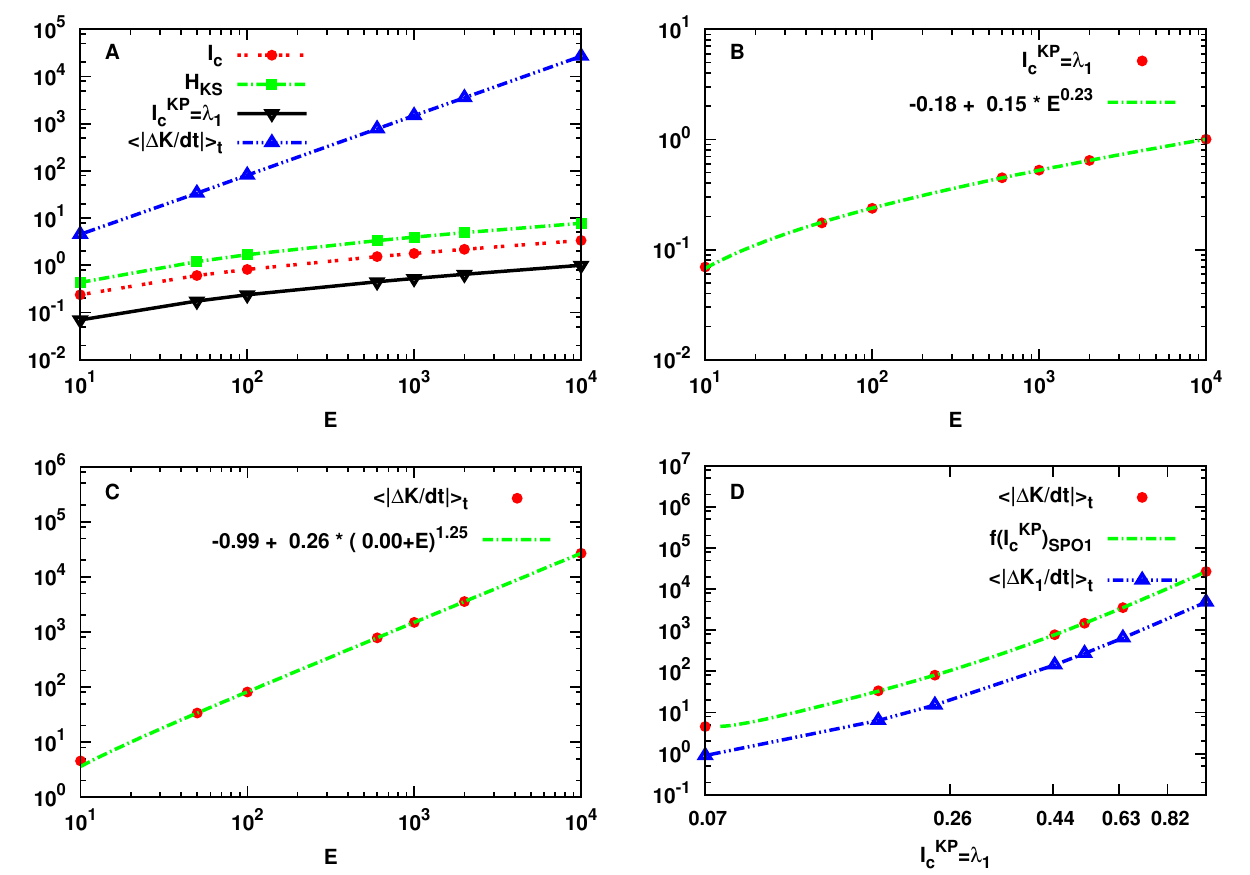}
\end{flushleft}
\caption{Panel A: Plot of quantities: $I_c$ as defined by Eq. \eqref{IcHamiltonian} in red dashed line with points, $H_{\mbox{KS}}$ as defined by Eq. \eqref{HKS_entropy} in green dashed line with rectangles, $I_{c}^{KP}$ as defined by Eq. \eqref{I_cKP} in black solid line with lower triangles and $\langle|\frac{\Delta K}{dt}|\rangle_t$ as defined by Eq. \eqref{averagetimeKdot} in blue dashed line with upper triangles as a function of $E$ for initial conditions $\vec{X}(0)$ located in the neighborhood of SPO1 of the FPU system. Note that both axes are logarithmic. Panel B: Plot of $I_c^{KP}=\lambda_1$ with red points with the power-law fitting of Eq. \eqref{LE1hamEfit_big_energy} in green line. Panel C: Plot of $\langle\bigl|\frac{\Delta K}{dt}\bigr|\rangle_t$ with red points with the power-law fitting of Eq. \eqref{timeaverageDKLE1fit_big_energy} in green line. Panel D: Power-law dependence of $\langle\bigl|\frac{\Delta K}{dt}\bigr|\rangle_t$ to $I_c^{KP}=\lambda_1$ in red points, in the interval $(0.07,1)$ that corresponds to the energy interval $[10,10^4]$ of panels A, B and C and of the power-law fitting of Eq. \eqref{timeaverageDKLE1fit_big_energy} in green dashed line. Note that all axes are logarithmic.
}
\label{IcLE_and_maxdK_dt_vs_nrg_COMBINED_ndof=15_big_nrg_interval_FPU_SPO1}
\end{figure}

In Fig. \ref{IcLE_and_maxdK_dt_vs_nrg_COMBINED_ndof=15_big_nrg_interval_FPU_SPO1} we present the corresponding plots and fits for the energy interval $[3,10^4]$ considering 15 particles and $\beta=1$. Following Ref. \cite{Antonopoulosetal2006a}, for these values we know that the dynamics around SPO1 is chaotic and thus allows the production and exchange of energy and information in the FPU chain.

\subsubsection{BEC OPM}\label{large_nrg_regime_BEC_OPM}

Next, we proceed and study the same problem for a different system, namely the BEC Hamiltonian given in Eq. \eqref{BEC_Hamiltonian_2}. We have chosen this system as it allows us to study the relation between transfer and exchange of energy and information in a different Hamiltonian system than the FPU. Furthermore, because it is not written in the form $H=K+P$ as the FPU does (compare Eqs. \eqref{FPU_Hamiltonian_beta} and \eqref{BEC_Hamiltonian_2}). It will thus permit us to demonstrate the validity of the upper bounds for the MIR and the connection between the exchange of energy and information in different observation spaces.

\begin{figure}[!ht]
\begin{flushleft}
\includegraphics[scale=1.2]{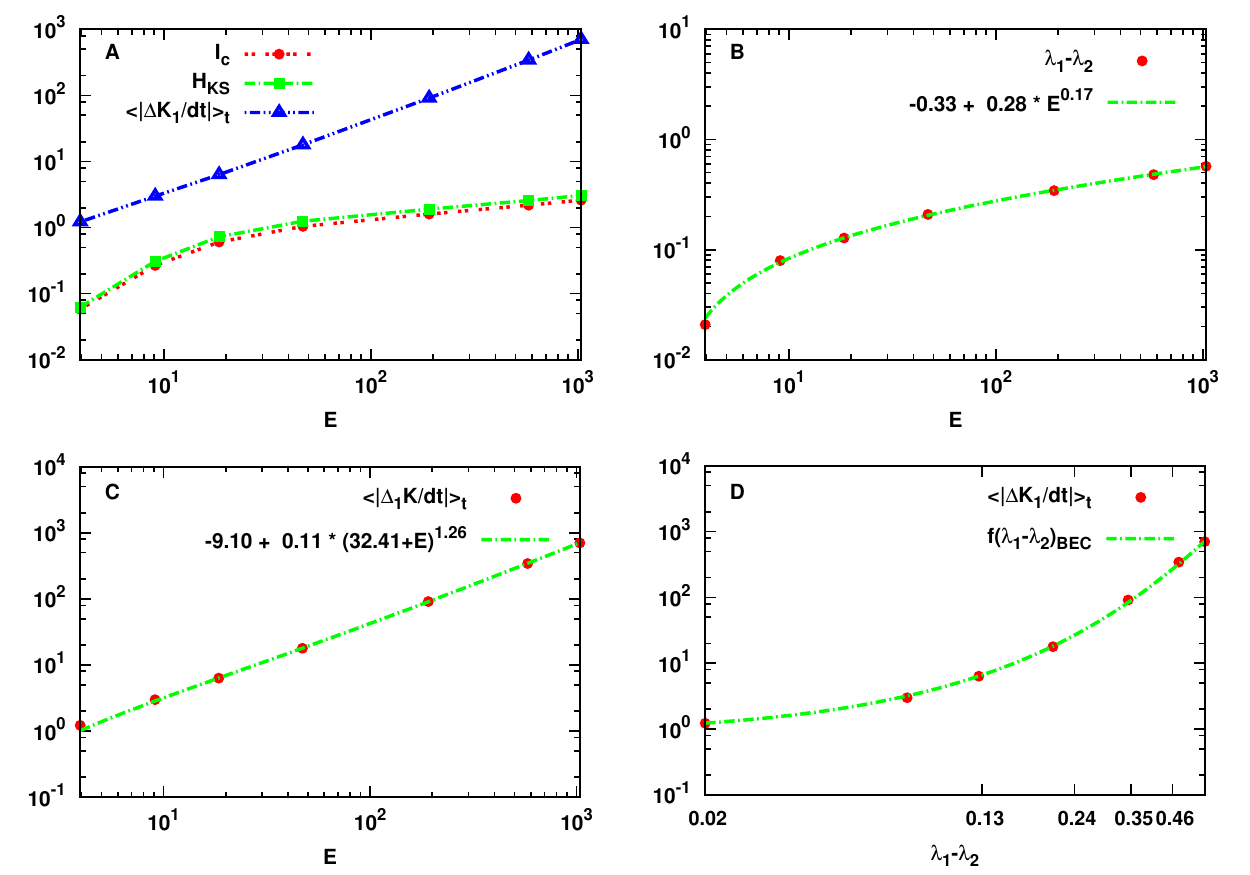}
\end{flushleft}
\caption{Panel A: Plot of quantities: $I_c$ as defined by Eq. \eqref{IcHamiltonian} in red dashed line with points, $H_{\mbox{KS}}$ as defined by Eq. \eqref{HKS_entropy} in green dashed line with rectangles, $I_{c}^{KP}$ as defined by Eq. \eqref{I_cKP} in black solid line with lower triangles and $\langle|\frac{\Delta K_1}{dt}|\rangle_t$ as defined by Eq. \eqref{Delta_K1} in blue dashed line with upper triangles as a function of $E$ for initial conditions $\vec{X}(0)$ located in the neighborhood of the OPM of the BEC Hamiltonian. Note that both axes are logarithmic. Panel B: Plot of $(\lambda_1-\lambda_2)$ with red points with the power-law fitting of Eq. \eqref{LE1hamEfit_big_energy} in green line. Panel C: Plot of $\langle\bigl|\frac{\Delta K_1}{dt}\bigr|\rangle_t$ with red points with the power-law fitting of Eq. \eqref{timeaverageDKLE1fit_big_energy} in green line. Panel D: Power-law dependence of $\langle\bigl|\frac{\Delta K_1}{dt}\bigr|\rangle_t$ to $(\lambda_1-\lambda_2)$ in red points, in the interval $(0.02,0.57)$ that corresponds to the energy interval $(3.94,1037.56)$ of panels A, B and C and of the power-law fitting of Eq. \eqref{timeaverageDKLE1fit_big_energy} in green dashed line. Note that all axes are logarithmic.
}
\label{IcLE_and_maxdK_dt_vs_nrg_COMBINED_ndof=6_big_nrg_interval_BEC_OPM}
\end{figure}

In particular, we consider here a small version of the system with $N=6$ degrees of freedom (particles) with initial conditions set in the neighborhood of the OPM periodic orbit given in Eq. \eqref{BEC_non_lin_mode_periodic_boundary_conditions_OPM} with periodic boundary conditions (see Eq. \eqref{BEC_periodic_boundary_conditions}). In Fig. \ref{IcLE_and_maxdK_dt_vs_nrg_COMBINED_ndof=6_big_nrg_interval_BEC_OPM}, we show the results of a similar study as we did in the cases of SPO1, SPO2 of the FPU system, for the energy interval $(3.94,1037.56)$ for which we have been able to study numerically in terms of the preservation of the accuracy of the computed energy. For this energy interval we know that the dynamics is chaotic (see Ref. \cite{Antonopoulosetal2006b}). Since, as we have already pointed out, BEC is not given by the sum of the kinetic and potential energy, we adopt a different strategy and reside on the calculation of the similar quantity $\langle\bigl|\frac{\Delta K_1}{dt}\bigr|\rangle_t$ based on the kinetic energy of the first particle $x_1$ (see Eq. \eqref{Delta_K1}). However, the kinetic energy of any other particle can be used as well. By fitting the data with the generalized power-laws of Eqs. \eqref{LE1hamEfit_big_energy}, \eqref{timeaverageDKLE1fit_big_energy} and \eqref{eq:main_result_1_big_nrg_interval} we have: $a_4\approx-0.33$, $b_4\approx0.28$, $c_4\approx0.17$ for Eq. \eqref{LE1hamEfit_big_energy}, $a_5\approx-9.1$, $b_5\approx0.11$, $c_5\approx32.41$, $d_5\approx1.26$  for Eq.\eqref{timeaverageDKLE1fit_big_energy} and finally: $a_6\approx-9.1$, $b_6\approx0.11$, $c_6\approx32.41$, $d_6\approx0.36$, $e_6\approx0.31$, $f_6\approx6.29$ and $g_6\approx1.27$ for Eq. \eqref{eq:main_result_1_big_nrg_interval}.

\section{Hamiltonian Communication System}\label{Ham_com_chaneel_FPU}

In this section we present an ``experimental'' implementation of a 1-dimensional communication channel based on the FPU Hamiltonian system of Eq. \eqref{FPU_Hamiltonian_beta}, and show the relation between our proposed upper bounds for the MIR with the actual MIR measured for the exchange of information between the first and last particle of the channel.

In more details, we consider the FPU chain of $N$ oscillators as a 1-dimensional communication channel where information and energy flow from one end to the other, i.e. from the first particle $x_1$ to the last one $x_N$ and vice versa. To extend the applicability of our theoretical results obtained in the previous sections for different cases of chaotic dynamics, we will use the dynamics around SPO1 and SPO2 and consider as a bi-dimensional observation space the one constructed by the evolution of the pair of position variables $x_1,x_N$ of the first and last particle of the FPU chain of Eq. \eqref{FPU_Hamiltonian_beta}. The computation of the actual MIR value between the two observation nodes $x_1$ ad $x_N$ was based on the theory presented in Ref. \cite{Baptistaetal2012}. Here, we consider 15 oscillators (degrees of freedom) for the SPO1 and 14 for the SPO2.

\begin{figure}[!ht]
\begin{flushleft}
\includegraphics[scale=1]{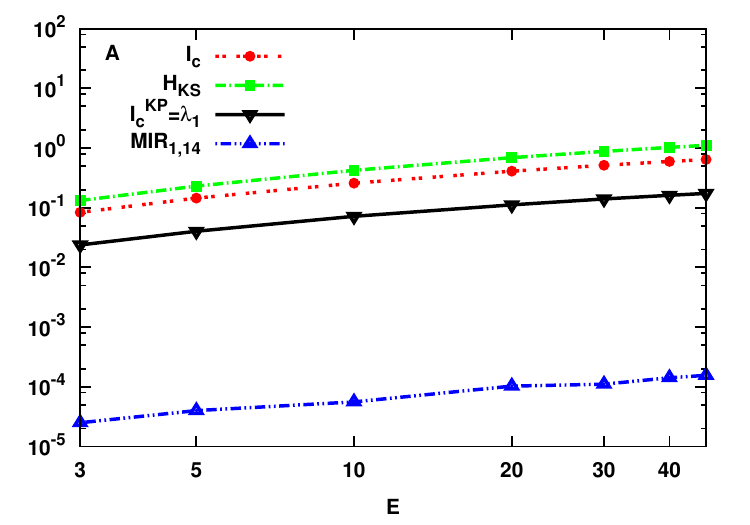}
\includegraphics[scale=1]{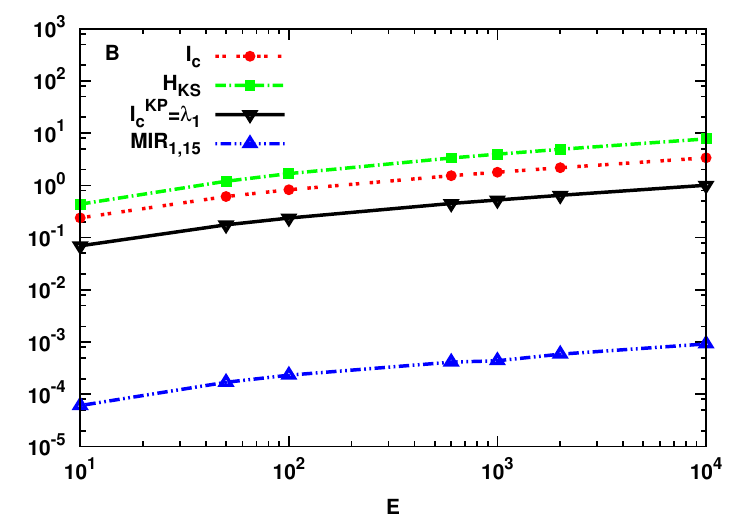}
\end{flushleft}
\caption{Panel A: Plot of the quantity $I_c$ as defined by Eq. \eqref{IcHamiltonian} in red dashed line with points, $H_{\mbox{KS}}$ as defined by Eq. \eqref{HKS_entropy} in green dashed line with rectangles, $I_{c}^{KP}$ as defined by Eq. \eqref{I_cKP} in black solid line with lower triangles and MIR$_{1,14}$ in blue dashed line with upper triangles as a function of $E$ for initial conditions $\vec{X}(0)$ set in the neighborhood of SPO2. Panel B: Same as in panel A for initial conditions set in the neighborhood of SPO1. Note that all axes are logarithmic.
}
\label{MIR_big_nrg_interval_FPU_SPO2_SPO1}
\end{figure}

In panel A of Fig. \ref{MIR_big_nrg_interval_FPU_SPO2_SPO1} we show the results of our study for the SPO2 case. We have plotted in red dashed line with points the quantity $I_c$ of Eq. \eqref{IcHamiltonian}, $H_{\mbox{KS}}$ as defined by Eq. \eqref{HKS_entropy} in green dashed line with rectangles, $I_{c}^{KP}$ of Eq. \eqref{I_cKP} in black solid line with lower triangles and MIR$_{1,14}$ in blue dashed line with upper triangles as a function of the energy $E$. Here, MIR$_{1,14}$ stands for the actual mutual information rate measured for the exchange of information between $x_1$ and $x_{14}$. From our theoretical results derived in the previous sections we expect that MIR$_{1,14}$ should be smaller or equal than $I_c^{KP}$. This is indeed what one observes as the MIR$_{1,14}$ curve is smaller than the previously mentioned upper bound and more importantly, it follows the same morphology (functional form) as $I_c$, $H_{KS}$ and $I_c^{KP}$. We have performed the same analysis for the SPO1 case as well showed in panel B of the same figure and arrive again at the same conclusions, i.e. MIR$_{1,15}$ lies below $I_c^{KP}$ as expected by our study and follows the same morphology as the upper bounds $I_c$, $H_{KS}$ and $I_c^{KP}$. In this case, MIR$_{1,15}$ denotes the actual mutual information rate measured for the exchange of information between $x_1$ and $x_{15}$. 

\section{Discussion}

In this paper we have studied the relation among the transfer of energy from kinetic ($K$) to potential ($P$) energies, the transfer of information between these two quantities and between different particles, the production of information, and Lyapunov exponents in Hamiltonian systems.

Our first result is that the largest Lyapunov exponent of the bi-dimensional space of the kinetic and potential energy is equal to the maximal Lyapunov exponent of the Hamiltonian in the case it is given by the sum of the kinetic and potential energy. Consequently, we were able to show that the upper bound for the MIR in the $KP$ subspace is given by the largest Lyapunov exponent of the Hamiltonian ($\lambda_1$). This implies that the more information the Hamiltonian system produces ($\lambda_1$), the more information can be exchanged between $K$ and $P$.

The second important result we have found is a power-law relation between the rate of transfer from kinetic to potential energy, the largest Lyapunov exponent of the Hamiltonian, and the Kolmogorov-Sinai entropy of the Hamiltonian. The more chaotic and the more information the Hamiltonian system produces ($\lambda_1$ and $H_{\mbox{KS}}$) respectively, the larger is the time average of the absolute value of energy transferred between $K$ and $P$ per unit of time (i.e. $\langle|\frac{\Delta K}{dt}|\rangle_t$).

The other important result is the proof of the inequality $I_c^{KP}<I_c<H_{\mbox{KS}}$ in the Appendix. It implies that, when one observes a Hamiltonian system through its kinetic and potential energies (thus obtaining $I_c^{KP}$), one measures less information about the Hamiltonian system than when observing half of its variables (thus obtaining $I_c$) or all of its variables (thus obtaining $H_{\mbox{KS}}$).

Finally, we have proposed an ``experimental'' implementation of a 1-dimensional communication channel based on a Hamiltonian system, and have calculated the actual rate with which information is exchanged between the first and last particle of the channel and compared that with the upper bounds we have proposed. As expected from our theoretical analysis, in all cases we have studied the actual MIR values were found to be smaller than our proposed upper bounds of MIR.

It is challenging to sketch here a possible connection between our results and the free energy $F$, entropy $S$, temperature $T$ and Hamiltonian energy $E$ in the thermodynamic limit, i.e. when $E$ and $N$ grow indefinitely while their ratio $E/N$ remains constant. According to the definition attributed to Helmholtz, $F$ is equal to the internal energy of the system $U$ minus the product of the (absolute) temperature $T$ multiplied by $S$, i.e. $F=U-TS$. $T$ is an important macroscopic quantity since its definition goes back to the early days of thermodynamics. Maxwell had realized that when the Hamiltonian has the special form:
\begin{equation*}
H(x,\dot{x})=\frac{1}{2}\dot{x}^{2}+P(x,\ldots),
\end{equation*}
(as is the case of the FPU system we have studied in this work) the canonical ensemble average of $\dot{x}^2$ is the temperature $T$ of the system. Thus, if one assumes ergodicity and equivalence of ensembles of initial conditions, it suffices to measure the time average of $\dot{x}^2$ during the evolution of the system in order to compute $T$ (see for example Ref. \cite{PhysRevLett.78.772}). Then, $U$ in this context is the fixed energy of the Hamiltonian (e.g. FPU) $E(=K+P)$ and $S$ can be calculated by the KS entropy $H_{\mbox{KS}}$ as $S=\alpha H_{\mbox{KS}}$, where $\alpha$ has the unit of time, since KS entropy is simply Shannon's entropy (equivalent to Gibb's entropy) per unit of time. Therefore, one can have:
\begin{eqnarray}
F&=&E-T\alpha H_{\mbox{KS}}\nonumber\Rightarrow\\
F&=&K+P-T\alpha H_{\mbox{KS}}\label{final_free_energy_eq},
\end{eqnarray}
and by solving Eq. \eqref{final_free_energy_eq} to obtain:
\begin{equation}\label{H_KS_free_energy}
 H_{\mbox{KS}}=\frac{K+P-F}{\alpha T}.
\end{equation}
If Eq. \eqref{eq:main_result_1} remains still valid in the thermodynamic limit, then by substituting Eq. \eqref{H_KS_free_energy} in the right hand side of Eq. \eqref{eq:main_result_1} one has:
\begin{equation}\label{our_results_and_free_energy}
	\langle\biggl|\frac{\Delta K}{dt}\biggr|\rangle_t\propto\Bigr({I_c^{KP}}\Bigl)^{\frac{b_2}{b_1}}=(\lambda_1)^{\frac{b_2}{b_1}}=\Biggr(2\frac{K+P-F}{\alpha TN}\Biggl)^{\frac{b_2}{b_1}},
\end{equation}
which relates the rate of transfer from kinetic to potential energy and the largest Lyapunov exponent of the Hamiltonian with the free energy and temperature of the system. This provides a direct relation between the results of this paper and important quantities of thermodynamics and Statistical Mechanics as long as the same conditions required for the derivation of the main results of our paper hold for Eq. \eqref{our_results_and_free_energy} as well. Equation \eqref{our_results_and_free_energy} implies that the larger the gap between the energy of the Hamiltonian and the available energy to do work (the free energy) the smaller the transfer of energy and information from $K$ to $P$ is.

In a series of papers \cite{Popovskietal2013,MohammadFouladgaretal2013,Toyabeetal2010,DBLP:journals/corr/abs-1303-1693}, the authors discuss about technological applications of the transfer of energy and information in communication, interference and graphical networks and show how one can reuse part of the energy for successive communication tasks. These ideas are based on results from physics showing that any system that exchanges information via the transfer of given physical resources such as radio waves, particles, etc., can reuse part of the received resources. If chaotic Hamiltonian systems could be used to create a communication system such that energy of the transmitting signal could be reused to transmit more information, from Eq. \eqref{our_results_and_free_energy} it is clear that $F$ must be different than zero implying that less information can be transmitted.

We believe that our work provides a viable pathway to establish similar relations between production and transfer of energy and information in other Hamiltonian systems for which the Lyapunov exponents have different dependences with the increase of the energy of the system as compared to those we have found here. Moreover, the choice of the bi-dimensional observation space is not restrictive and a plausible one can be constructed by the position coordinates of any two particles of the system. Of course, in these cases it is expected that our power-law relations will be replaced by new ones reflecting the different properties of the systems.

\section*{Appendix}

Here, we prove the other main result of this paper which is the inequality:
\begin{equation}\label{second_result}
I_c^{KP}<I_c<H_{\mbox{KS}} 
\end{equation}
and thus explain the result of Fig. \ref{IcLE_and_maxdK_dt_vs_nrg_Ham}. In Ref. \cite{Livietal1986}, the authors discuss about the existence of the spectrum of the Lyapunov exponents in the thermodynamic limit and investigate numerically this existence in the FPU-$\beta$ model given herein by Eq. \eqref{FPU_Hamiltonian_beta}. They show that the shape of the Lyapunov spectrum for energy densities $\epsilon=E/N$ well above the equipartition threshold allows someone to express $H_{\mbox{KS}}$ in terms of the largest Lyapunov exponent $\lambda_1$ only:
\begin{equation}\label{HKSintegral}
	H_{\mbox{KS}}=\int_0^{\lambda_1}\lambda C N d\lambda=\frac{N}{2}\lambda_1,
\end{equation}
where $C=1/\lambda_1$.

By applying the above ideas in our case for $E\in(E_{\mbox{u}},E_{\mbox{r}})$ and using Eq. \eqref{HKSintegral} we have:
\begin{eqnarray}
	H_{\mbox{KS}}&=&\int_0^{\lambda_1}\lambda C N d\lambda\label{HKSHamiltonian_integral}\Rightarrow\\
	H_{\mbox{KS}}&=&\int_0^{\lambda_{N/2}}\lambda C N d\lambda + \int_{\lambda_{N/2}}^{\lambda_1}\lambda C N d\lambda\nonumber\Rightarrow\\
	H_{\mbox{KS}}&=&\frac{N}{2\lambda_1} \bigr(\lambda^2\bigl)_0^{\lambda_{N/2}} + \tilde{H}\nonumber\Rightarrow\\
	\tilde{H}&=&H_{\mbox{KS}}-\frac{N}{2\lambda_1}\bigl(\lambda_{N/2}\bigr)^2\label{Htilde},
\end{eqnarray}
where we have used $C=1/\lambda_1$ and $\bar{H}=\int_{\lambda_{N/2}}^{\lambda_1}\lambda C N d\lambda$. Term $\lambda_{N/2}$ is the $(N/2)$th positive Lyapunov exponent of Hamiltonian \eqref{FPU_Hamiltonian_beta} when sorting them in descending order (i.e. $\lambda_1>\lambda_2>\ldots>\lambda_{N/2}>\ldots>\lambda_N=0$). It comes from the fact that in Eq. \eqref{HKSHamiltonian_integral} we integrate over all positive Lyapunov exponents and that we want to relate $H_{\mbox{KS}}$ with $\tilde{H}$ of Eq. \eqref{IcHamiltonian} which is defined as the sum over the first $N/2$ positive Lyapunov exponents when they are sorted in descending order.

By substituting Eq. \eqref{Htilde} in Eq. \eqref{IcHamiltonian} we have:
\begin{eqnarray}
	I_c&=&2\tilde{H}-H_{\mbox{KS}}\nonumber\Rightarrow\\
	I_c&=&H_{\mbox{KS}}-\frac{N}{\lambda_1}\bigl(\lambda_{N/2}\bigr)^2\label{Icham},
\end{eqnarray}
and so we obtain:
\begin{equation}
	I_c<H_{\mbox{KS}}\label{IcHam2stHKSHam}
\end{equation}
which is the right hand side inequality of Eq. \eqref{second_result}.

By combining Eqs. \eqref{HKSintegral}, \eqref{Icham} and setting $I_c^{KP}=\lambda_1$, we obtain:
\begin{eqnarray}
	I_c&=&\frac{N}{2}\lambda_1-\frac{N}{\lambda_1}\bigl(\lambda_{N/2}\bigr)^2\nonumber\Rightarrow\\
	I_c&=&\frac{N}{2}I_c^{KP}-\frac{N}{I_c^{KP}}\bigl(\lambda_{N/2}\bigr)^2\label{Icham2}.
\end{eqnarray}
The last equation links the upper bound of information transfer in the phase space of the Hamiltonian with the upper bound of the information that can be transferred in the $KP$ space. Moreover, an important consequence of Eq. \eqref{Icham} is that $I_c=H_{\mbox{KS}}$ when $\lambda_{N/2}=0$ implying that this can happen when there are at least $N/2$ integrals of motion and leading to the conclusion that it should be $\lambda_{N/2}=\lambda_{(N/2)+1}=\ldots=\lambda_{N}=0$. However, this is not happening in our case since all Lyapunov exponents are positive but the last one $\lambda_{N}=0$ as the Hamiltonian is an integral of the motion.

Next, we prove the left hand side inequality of Eq. \eqref{second_result}:
\begin{equation}
  I_c^{KP}<I_c\label{I_c_tmp1}.
\end{equation}
To do so, let us suppose that:
\begin{equation}
	I_c-I_c^{KP}=0\label{IcHamgtIcKP}
\end{equation}
and check under which assumptions for $I_c^{KP}$ Eq. \eqref{I_c_tmp1} holds.
For this, we substitute Eq. \eqref{Icham2} for $I_c$ into Eq. \eqref{IcHamgtIcKP} and have:
\begin{equation}
	\biggl(\frac{N-2}{2}\biggr)\bigl(I_c^{KP}\bigr)^2-N\bigl(\lambda_{N/2}\bigr)^2=0\label{seconddegreepolynom}.
\end{equation}
The last equation is a second degree polynomial with respect to $I_c^{KP}$. Its determinant is given by:
\begin{equation}\nonumber
	\mathcal{D}=2N(N-2)\bigl(\lambda_{N/2}\bigr)^2,
\end{equation}
which is positive for $N>2$ and thus, the two discrete real roots are:
\begin{eqnarray}
	I_c^{KP}&=&\frac{\lambda_{N/2}\sqrt{2N(N-2)}}{N-2}>0\mbox{ and}\label{root1}\\
	I_c^{KP}&=&-\frac{\lambda_{N/2}\sqrt{2N(N-2)}}{N-2}<0\nonumber.
\end{eqnarray}
By theory, we know that Eq. \eqref{seconddegreepolynom} is positive and thus inequality in Eq. \eqref{I_c_tmp1} is true when $I_c^{KP}>\frac{\lambda_{N/2}\sqrt{2N(N-2)}}{N-2}$ since the term $\frac{N-2}{2}$ of $I_c^{KP}$ is positive for $N>2$.

The second root is not physically possible to exist since it would imply that $I_c^{KP}<0$ for $N>3$ contradicting to the fact that $I_c^{KP}$ is positively defined. Thus, Eq. \eqref{IcHamgtIcKP} is positive when $I_c^{KP}>\frac{\lambda_{N/2}\sqrt{2N(N-2)}}{N-2}$, which is always true, since $\lambda_{N/2}\ll1$ and:
\begin{eqnarray}\nonumber
	\lim_{N\rightarrow\infty}\frac{\sqrt{2N(N-2)}}{N-2}=\sqrt{2}.
\end{eqnarray}
Thus, we have proved that:
\begin{equation}
	I_c^{KP}<I_c\label{IcHam2gtIcKP}.
\end{equation}
Combining Eqs. \eqref{IcHam2stHKSHam} and \eqref{IcHam2gtIcKP}, we obtain:
\begin{equation}\label{3part_inequality}
	I_c^{KP}<I_c<H_{\mbox{KS}}.
\end{equation}
The way $I_c$ is defined (see Eq. \eqref{IcHamiltonian}) implies that $I_c<H_{\mbox{KS}}$ since $\tilde{H}<H_{\mbox{KS}}$. In panel A of Fig. \ref{IcLE_and_maxdK_dt_vs_nrg_Ham} we can check that indeed inequality \eqref{3part_inequality} is fulfilled.

Finally, it worths mentioning that according to Eq. \eqref{root1} it is possible to have:
\begin{equation}\nonumber
	I_{c}^{KP}=I_c
\end{equation}
that is, the upper bounds of information transfer in the bi-dimensional subspace and in the Hamiltonian to be equal when it happens that:
\begin{equation}\nonumber
	I_c^{KP}=\frac{\lambda_{N/2}\sqrt{2N(N-2)}}{N-2}=\sqrt{2}\lambda_{N/2}.
\end{equation}
The last equation provides an alternative estimation of $I_c^{KP}$ valid when:
\begin{equation}\nonumber
	I_c^{KP}=\lambda_1=\sqrt{2}\lambda_{N/2}.
\end{equation}

\section*{Acknowledgments}

Ch. A. would like to dedicate this paper to his recently born nephew, sharing both a common period of being ``in preparation''. The authors were partially supported by the ``EPSRC EP/I032606/1'' grant.


\section*{Figure Legends}
\setcounter{figure}{0}

\begin{figure}[!ht]
\caption{{\bf Schematic representation of the time evolution after one time step $dt$ of two deviation vectors (arrows) corresponding to the direction along the Lyapunov exponent $\lambda_1^{KP}$ on the 1-dimensional subspace $K+P=E$ on the $KP$ space}. $\vec{X_1}(t)$ and $\vec{X_2}(t)$ are two trajectories in the phase space of Hamiltonian \eqref{FPU_Hamiltonian_beta} that drive the dynamics along this line. We denote with $\delta$ and $\Delta$ the lengths of the two deviation vectors initially and after one time step, respectively.}
\end{figure}

\begin{figure}[!ht]
\caption{{\bf Plot of the absolute difference $|\lambda_1-\lambda_1^{KP}|$ as a function of time for two trajectories $\vec{X_1}(t)$ and $\vec{X_2}(t)$ located initially in the neighborhood of SPO2 at the same energy $E$}. Here, $E=30$ is well inside the interval $(E_{\mbox{u}},E_{\mbox{r}})$. Note that both axes are logarithmic.}
\end{figure}

\begin{figure}[!ht]
\caption{{\bf Plot of the quantities: $I_c$ as defined by Eq. \eqref{IcHamiltonian} in red dashed line with points, $H_{\mbox{KS}}$ as defined by Eq. \eqref{HKS_entropy} in green dashed line with rectangles, $I_{c}^{KP}$ as defined by Eq. \eqref{I_cKP} in black solid line with lower triangles and $\langle|\frac{\Delta K}{dt}|\rangle_t$ as defined by Eq. \eqref{averagetimeKdot} in blue dashed line with upper triangles as a function of $E$ for initial conditions $\vec{X}(0)$ located in the neighborhood of SPO2 of the FPU system}. Note that both axes are logarithmic.}
\end{figure}

\begin{figure}[!ht]
\caption{{\bf Panel A: Plot of quantities: $I_c$ of Eq. \eqref{IcHamiltonian} in red dashed line with points and $H_{\mbox{KS}}$ of Eq. \eqref{HKS_entropy} in green dashed line with rectangles}. Panel B: Plot of quantities $I_c^{KP}=\lambda_1$ with red points with the power-law fitting of Eq. \eqref{LE1hamEfit} in green line. Panel C: Plot of $\langle\bigl|\frac{\Delta K}{dt}\bigr|\rangle_t$ with red points with the power-law fitting of Eq. \eqref{timeaverageDKLE1fit} in green line. Panel D: Power-law dependence of $\langle\bigl|\frac{\Delta K}{dt}\bigr|\rangle_t$ to $I_c^{KP}=\lambda_1$ in red points, in the interval $(0.140,0.174)$ that corresponds to the energy interval $[30,47]$ of panels A, B and C and of the power-law fitting of Eq. \eqref{timeaverageDKLE1fit} in green dashed line. Note that all axes are logarithmic.}
\end{figure}

\begin{figure}[!ht]
\caption{{\bf Panel A: Plot of quantities: $I_c$ as defined by Eq. \eqref{IcHamiltonian} in red dashed line with points, $H_{\mbox{KS}}$ as defined by Eq. \eqref{HKS_entropy} in green dashed line with rectangles, $I_{c}^{KP}$ as defined by Eq. \eqref{I_cKP} in black solid line with lower triangles and $\langle|\frac{\Delta K}{dt}|\rangle_t$ as defined by Eq. \eqref{averagetimeKdot} in blue dashed line with upper triangles as a function of $E$ for initial conditions $\vec{X}(0)$ located in the neighborhood of SPO2 of the FPU system}. Note that both axes are logarithmic. Panel B: Plot of $I_c^{KP}=\lambda_1$ with red points with the power-law fitting of Eq. \eqref{LE1hamEfit_big_energy} in green line. Panel C: Plot of $\langle\bigl|\frac{\Delta K}{dt}\bigr|\rangle_t$ with red points with the power-law fitting of Eq. \eqref{timeaverageDKLE1fit_big_energy} in green line. Panel D: Power-law dependence of $\langle\bigl|\frac{\Delta K}{dt}\bigr|\rangle_t$ to $I_c^{KP}=\lambda_1$ in red points, in the interval $(0.02,0.174)$ that corresponds to the energy interval $[3,47]$ of panels A, B and C and of the power-law fitting of Eq. \eqref{timeaverageDKLE1fit_big_energy} in green dashed line. Note that all axes are logarithmic.
}
\end{figure}

\begin{figure}[!ht]
\caption{{\bf Panel A: Plot of quantities: $I_c$ as defined by Eq. \eqref{IcHamiltonian} in red dashed line with points, $H_{\mbox{KS}}$ as defined by Eq. \eqref{HKS_entropy} in green dashed line with rectangles, $I_{c}^{KP}$ as defined by Eq. \eqref{I_cKP} in black solid line with lower triangles and $\langle|\frac{\Delta K}{dt}|\rangle_t$ as defined by Eq. \eqref{averagetimeKdot} in blue dashed line with upper triangles as a function of $E$ for initial conditions $\vec{X}(0)$ located in the neighborhood of SPO1 of the FPU system}. Note that both axes are logarithmic. Panel B: Plot of $I_c^{KP}=\lambda_1$ with red points with the power-law fitting of Eq. \eqref{LE1hamEfit_big_energy} in green line. Panel C: Plot of $\langle\bigl|\frac{\Delta K}{dt}\bigr|\rangle_t$ with red points with the power-law fitting of Eq. \eqref{timeaverageDKLE1fit_big_energy} in green line. Panel D: Power-law dependence of $\langle\bigl|\frac{\Delta K}{dt}\bigr|\rangle_t$ to $I_c^{KP}=\lambda_1$ in red points, in the interval $(0.07,1)$ that corresponds to the energy interval $[10,10^4]$ of panels A, B and C and of the power-law fitting of Eq. \eqref{timeaverageDKLE1fit_big_energy} in green dashed line. Note that all axes are logarithmic.
}
\end{figure}

\begin{figure}[!ht]
\caption{{\bf Panel A: Plot of quantities: $I_c$ as defined by Eq. \eqref{IcHamiltonian} in red dashed line with points, $H_{\mbox{KS}}$ as defined by Eq. \eqref{HKS_entropy} in green dashed line with rectangles, $I_{c}^{KP}$ as defined by Eq. \eqref{I_cKP} in black solid line with lower triangles and $\langle|\frac{\Delta K_1}{dt}|\rangle_t$ as defined by Eq. \eqref{Delta_K1} in blue dashed line with upper triangles as a function of $E$ for initial conditions $\vec{X}(0)$ located in the neighborhood of the OPM of the BEC Hamiltonian}. Note that both axes are logarithmic. Panel B: Plot of $(\lambda_1-\lambda_2)$ with red points with the power-law fitting of Eq. \eqref{LE1hamEfit_big_energy} in green line. Panel C: Plot of $\langle\bigl|\frac{\Delta K_1}{dt}\bigr|\rangle_t$ with red points with the power-law fitting of Eq. \eqref{timeaverageDKLE1fit_big_energy} in green line. Panel D: Power-law dependence of $\langle\bigl|\frac{\Delta K_1}{dt}\bigr|\rangle_t$ to $(\lambda_1-\lambda_2)$ in red points, in the interval $(0.02,0.57)$ that corresponds to the energy interval $(3.94,1037.56)$ of panels A, B and C and of the power-law fitting of Eq. \eqref{timeaverageDKLE1fit_big_energy} in green dashed line. Note that all axes are logarithmic.
}
\end{figure}

\begin{figure}[!ht]
\caption{{\bf Panel A: Plot of the quantity $I_c$ as defined by Eq. \eqref{IcHamiltonian} in red dashed line with points, $H_{\mbox{KS}}$ as defined by Eq. \eqref{HKS_entropy} in green dashed line with rectangles, $I_{c}^{KP}$ as defined by Eq. \eqref{I_cKP} in black solid line with lower triangles and MIR$_{1,14}$ in blue dashed line with upper triangles as a function of $E$ for initial conditions $\vec{X}(0)$ set in the neighborhood of SPO2}. Panel B: Same as in panel A for initial conditions set in the neighborhood of SPO1. Note that all axes are logarithmic.
}
\end{figure}

\section*{Tables}

\end{document}